# Quantum action, non-locality and coherence from classical perception – a new facet of Lagrangian formalism for relativistic dynamics

D. Das (dasd1951@gmail.com)
(Retired from Bhabha Atomic Research Centre
Trombay, Mumbai 400 08)

Classical and quantum mechanical descriptions of physical world are seamlessly abridged within the framework of Lagrangian formalism which, besides revealing the essence of nonlocally correlated dynamic evolution, helps understanding abrupt onset towards perturbation driven correlation breakdown criticality with the manifestation of classical dynamic properties. The abridged formalism takes into consideration a family of ubiquitously correlated paths in their linear combination and shows that the coherently evolved dynamic course of optimum displacement is variationally realizable within arbitrarily selected pair of space-like surface-boundaries, only if the nonlocal correlation implies virtually mediating energy and momentum quanta across instantly evolved volume meeting integral conservations. The coherent evolution characteristics are explored for correlated geodesics of quantum space and for dynamics of particle. As safeguard to coherently mediated system, ubiquitous nonlocal mitigation on external perturbation is explored in explaining decoherence criticality considering the case of electrodynamic evolution.

Coherently evolved geodesics endorse quantum space of ultrahigh energy harmonic oscillators, which is exotic to the physical world described by evolution harmonics of comparatively low energies over a wide frequency range. Observable recessional kinematics of cosmic space with its background vacuum field supporting the physical world is analyzed by considering nonlocal mitigation of exotic QV field on the gravitationally perturbed space across ubiquitous isolation barrier arising from markedly different quantum orders of the two vacuum fields describable by a common dimensional element like Planck’s length. Analysis includes mitigating interaction of exotic QV field involving quantum tunneling across the energy barrier leading to the space recession kinematics. It helps explaining presence of the invisible signatures of dark matter and energy components with their relative proportions and also in rationalizing ubiquitously maintained causality in all cosmic phenomena.

.

## Introduction

Historically, the quantum mechanics explored the importance of nonlocal correlation in the analysis of dynamic evolutions of physical world. Nonlocally correlated evolution established through axiomatic introduction of quantum action within the framework of classical formalism conforms to the delocalized description of canonically correlated dynamic properties meeting conservation of quantized energy and momentum integrally in instant evolution volume. In a coherent dualism of matter and wave properties the quantum formalism works out the nonlocally governed existential probabilities of a particle in evolution volume at any instant. Probabilistic existence within instant volume is normalized to unity to endorse overall integrity as a particle in the delocalized description. For quantum mechanically entangled particles, the 'instant correlation' prevails among the coherently evolving partners [1,2,3,4]. Driven by perturbation of external fields, evolution harmonics of the nonlocally correlated quantum system can be critically over-beaten to result in breakdown of coherency in evolution with the consequential power relaxation with the revelation of classical kinematics. Thus, by interception of measuring (macroscopic) apparatus, the delocalized description abruptly turns into perceptibly localized one with apparent loss of the nonlocally mediated unitary evolution maintaining causality principle. The observationally based switchover mechanism to classically evolved state is apparently instantaneous as has been inferred from 'wave vector collapse' of widely separated entangled pair of particles/photons [5,6]. The stated weakness of abrupt transit from nonlocal to locally described evolution calls for rethinking beyond the axiomatically achieved formalism. Past attempts in bridging the gap between classical and quantum formalisms considered the dynamic action as pertinent property in understanding the nonlocal influence in field-particle interaction. Thus, the nonlocal influence has been addressed in variant forms, such as the quantum potential [7] in correlating classical trajectories prescribed by Hamilton-Jacobi equation, or, as ubiquitously mediated diffusion transport of quantized energy-momentum implied by Schrodinger wave evolution [8], or, as in Feynman's approach of the sum-over-(path) histories in predicting the probabilistic arrival of a particle at a space-time position from a source, which considers quantized phase factor in expressing superimposed amplitudes of matter waves propagating out from the past position [9]. In these attempts, the abrupt switchover property of nonlocal governance with the inception of external perturbation on coherently evolved state remains unanswered. Abrupt switching over of nonlocal correlation to locally governed one defines the doorstep of quantum to classical regime.

Quantum mechanics refers space-time solely for the purpose of describing evolutions of physical systems in coordinate space, (alternatively, in the reciprocal space) and by that it invites interpretational problems on several counts within the doctrine of ubiquitous non-locality involved in the axiomatic formalism: the wave-particle duality, the probabilistic existence of particle, and the quantum statistics of identical particles. Quantum field theory that uniquely takes care of these problems invokes incessant creations and annihilations of particles and their antiparticles in pairs within the fluctuating background vacuum field of cosmic space. The QM uncertainty principle is thereby rightfully placed within the framework of QFT that refers to the fluctuating background field as real vacuum of gravitating cosmic space with its zero point energy field composed of harmonic oscillator modes of all frequencies. Fluctuation refers to the variance of field strength in the minimal energy state of background field of real vacuum, which is inseparably involved in all QFT analyses that stood tests of its several predictions, particularly in quantum electrodynamics. The predicted anomalous magnetic moment of electron is one such example of the QFT analysis. As the consequence of zero point energy of fluctuating field, the observed Casimir effect adds testimony to quantum ordered cosmic background of real vacuum field supporting the physical world. The stable/quasi-stable existences of all physical systems in space with their evolution harmonics are in a way approved by the fluctuating background vacuum field with oscillator modes of all frequencies.

Amidst all the positive features, QFT formalism encounters an outstanding issue in reasoning out the observable recession property of the gravitationally active cosmic space. Observably, the flat space with near homogenous distribution of gravitating matters in cosmic scale is undergoing speedy recession [10] in isotropic manner. The observation calls for the involvement of invisible repulsive field that uniformly reacts everywhere surpassing the inward pull of gravitating matters. QFT reinterpreted the once considered Lamda term of energy-momentum field equations of general relativity as accounting for the contribution from repulsive internal pressure corresponding to the zero point energy density of fluctuating background vacuum field. Repulsive property of the vacuum field energy is anticipated from thermodynamic argument on the equation of state with internal energy density, $\rho_{vac}^{0}$; the repulsive internal pressure being expressible as, $p_{vac}^{0} = -\rho_{vac}^{0}$. However, the reasoning encounters tremendous setback from the utterly high value of quantum mechanically predicted energy density (~$10^{114}$Jm$^{-3}$) as compared to the critical density (~8x$10^{-10}$Jm$^{-3}$) of the observably flat space. The QFT predicted energy density exceeds the critical density of cosmic space by a factor of several dozens of orders of magnitude. The predicted vacuum property is based on the consideration of smallest dimension of space ($l_p$) and time ($t_p$) as was empirically proposed by Max Planck through the consolidations of three fundamental constants as $l_p = \sqrt{G\hbar / c^3}$ and $t_p = l_p / c$, where $G$, $\hbar$, and $c$ represent gravitation constant, reduced action constant, and signal speed in vacuum respectively. Vacuum field conceivable thereof belongs to a family of quantum oscillators of ultrahigh energies (at and above, $2\pi c\hbar / l_p$, $l_p$~1.616x$10^{-35}$m). The gravitating cosmic space with the billions of year long recessional history of physical world is not expected to coexist with the ubiquitous quantum vacuum possessing exotic field of ultrahigh energies. The stated setback in accounting for the repulsive influence in cosmic space recession suggests the need of understanding whether QM predicted exotic field can directly interact with cosmic space at all. The background vacuum field (BV field) of cosmic space supporting the physical world in entirety involves evolution harmonics of a wide range of frequencies from very low to very high values. It resonantly/quasi-resonantly supports fundamental particles of low to very high masses. The background vacuum as ideally extrapolated matter-free state of the gravitating cosmic space apparently belongs to a lattice ordered quantum state made of coherent dimension that supports the resonance lengths of subatomic particles ($l_R = 2\pi c\hbar / E_R$, $E_R$ being resonance energy). Resonance length can be as low as that of heaviest known boson, namely, Higgs particle with energy of about 125 GeV ($l_R$ ~$10^{-17}$m). Coherent length of the quantum space is expected to be of lower orders as compared to $10^{-17}$m because of the fact that for the stable/quasi stable existence of subatomic particles in space, their resonance lengths closely fits with integral multiples of the basic space dimension. With the assumption that the basic quantum dimension is unique irrespective of the two types of vacuum spaces, namely, the exotic QV with field having wave dispersion characteristics, $\omega \geq 2\pi c / l_p$, and the background vacuum of cosmic space supporting evolution harmonics of a wide range, $0 \leq \omega \leq 2\pi c / l_p$, ($2\pi c / l_p$ being pragmatically considered highest frequency limit), one finds that the two vacuum spaces possess strikingly different orders of their evolution harmonics. Oscillator harmonics of mighty QV field evolve without any specific order among the quantum mechanically allowed oscillator modes, whereas the oscillator harmonics of BV field are apparently in conformation with the lattice-like virtual ordering. In this study, the coexistence and interaction properties of ubiquitous QV field with the cosmic BV field of distinguishable quantum order are analyzed for understanding the observable recession kinematics of space. For the analysis, evolution properties of exotic vacuum field is worked out at first by developing an ab-initio formalism of the nonlocally correlated evolution harmonics of geodesics of 4-space in general. A basic quantum length governing wave dispersion characteristics of the concerned evolution harmonics is then evaluated instead of using the empirically proposed basic space dimension, $l_p$. Arrived quantum length deciding the coherent dimension of quantum space is then used to understand the virtual lattice ordered state of real (background) vacuum of cosmic space as compared to the coherent length associated with exotic QV field described by harmonic modes.

The ab-initio analysis used in the description of nonlocally correlated geodesics of quantum space has been extended to explore the ubiquitous involvement of nonlocal influence on a coherently evolved dynamic system. The exploration helps obtaining the evolved properties in accordance with the quantum mechanically formulated known results. In addition, with the inception of an external perturbation the exploration helps analyzing ubiquitous safeguard characteristics of nonlocally mediated evolution. These are exemplified by taking the case of dynamic evolution of a point-like charge in the presence of electromagnetic field. Coherently mediated dynamics when driven by externally governed field, the onset of decoherence criticality under defense reaction from ubiquitously mitigating safeguard and also the maximum accessibility of nonlocal defense at the criticality are addressed in understanding the brief onset course of mitigated evolution which culminates with the perturbation driven wave vector collapse at radiation boundary of evolved charge. Thus, the classically based Lorentz-Abraham-Dirac's radiation 4-force of electrodynamics has been rationalized as the causal consequence of the brief onset course of externally driven dynamic evolution culminating with the radiation relaxation and reaction thereof from the momentum loss. Generally, it is noted that in nonlocally mediated evolution the ubiquitous mitigation aspect on external perturbation leading to decoherence criticality are having features which are irrespective of whether the evolution concerns perturbed dynamics of interacting field and particle, or, concerns gravitationally perturbed cosmic space. Causality is globally maintained in coherent evolution and in its externally perturbed state culminating with the abrupt loss of coherence through relaxations with changeover to classical evolution. Overall it is shown that the ubiquitous mediation and mitigation maintaining causality in all dynamic evolutions are the outcome of mighty quantum vacuum that relentlessly governs the gravitationally perturbed cosmic space indirectly across a high kinematic barrier ever since the observable space came into existence. Part-A of the text deals with nonlocally correlated evolution properties of geodesics including the quantum features of space in the presence/absence of gravitating matters. Part-A also includes the relevance of mighty quantum vacuum field in gravitating cosmos with its fluctuating background vacuum field undergoing recession with the observable characteristics. As complement to the analysis presented in Part-A, Part-B of the text exemplifies nonlocal mediation features and also, the mitigating safeguard characteristics of the ubiquitous governance in coherently evolved dynamics.

## PART-A

### A 1.0 **Nonlocally entangled geodesics of quantized space**

In dynamic analysis, space is referred with its geodesic field as governed by the evolution property of space metric. Classically described geodesics need reform when metric evolution is ruled by the ubiquitous nonlocal mediation in quantum realm. Required step in obtaining reformed space evolution property would be the endowment of optimum displacement criterion in nonlocally correlated geodesics network considered in between arbitrarily selected pair of space-like surfaces having time-like separation.

Nonlocally mediated evolution property of geodesics in 4-space is thus arrived by considering a family of paths $\left\{x_p^{\alpha}(\tau)\right\}$, which in their instantly correlated linear combinations meet the optimum displacement criterion in the evolution course in space. For the nonlocally correlated displacement course the arbitrarily selected pair of space-like surfaces are referred by the boundary values of time-like parameter, $\tau$. Thus, the optimally displaced course of the nonlocal correlated paths can be realized through variational optimization as expressed by $\delta\int_{\tau_1}^{\tau_2}\left(\sum_p c_p^2 g_{\alpha\beta} dx_p^\alpha dx_p^\beta\right)^{1/2} = 0$, and the expression is valid irrespective of the boundary limits, $\tau_1$ and $\tau_2$ including the limiting case of convergence, $\tau_1 \to \tau_2$. Under the integral differential argument, $\left(\sum_p c_p^2 g_{\alpha\beta} dx_p^\alpha dx_p^\beta\right)^{1/2} = d\tau$ (say) represents linearly combined elemental displacement, where $g^{\alpha\beta}(x)$ is symmetric metric of space having signature (1,-1,-1,-1); path family $\{x_p^\sigma\}$, ($\sigma = 0,1,2,3$), belonging to 4-space ($x$). Weight factors $\{c_p\}$ associated with the linearly combined displacement components, $\{dx_p^\alpha\}$, as $dx^\alpha = \sum_p c_p dx_p^\alpha$, conform to the instant correlation characteristics, $c_p c_{p'} = \delta_{pp'}$, ($\delta_{pp'} = 0$ for $p \neq p'$, and $\delta_{pp'} = 1$ for $p = p'$); weight factors $\{c_p\}$ corroborating to the normalization $\sum_p c_p^2 = 1$. In the absence of external influence, the instantly referred weight factors are nonlocally governed and do not bear externally driven stochastic components. The variational optimization essentially concerns with the optimum magnitude of linearly combined displacement in between the arbitrarily selected variation boundaries at $\tau_1$ and $\tau_2$. Characteristically, the correlated displacement of path family is controlled by ubiquitous interplay of nonlocal kind implied by the involvement of canonically attributed weight factors. The displacement criterion applied to the time evolved course leads to the following equation:

$$(v_\beta \delta x^\beta)\Big|_{\tau_1}^{\tau_2} - \int_{\tau_1}^{\tau_2} \delta x^\beta [\dot{v}_\beta - v^\alpha v^\sigma \partial_\beta g_{\alpha\sigma}/2] d\tau = 0 \qquad (1),$$

where, $\delta x^\alpha \equiv \sum_p c_p \delta x_p^\alpha$, $v^\alpha(\tau) \equiv dx^\alpha/d\tau = \sum_p c_p v_p^\alpha(\tau)$, $\dot{v}^\alpha(\tau) \equiv d^2x^\alpha/d\tau^2 = \sum_p c_p \dot{v}_p^\alpha(\tau)$, and $\partial_\beta g_{\alpha\sigma} \equiv \partial g_{\alpha\sigma}/\partial x^\beta$, ($c_p c_{p'} = \delta_{pp'}$, $\sum_p c_p^2 = 1$). Displacement rate, $v^\alpha(\tau)$ corresponds to canonical representation of instant 4-velocity and accordingly, $\dot{v}^\alpha(\tau)$ is 4-acceleration at that instant. In their covariant forms, the displacement related 4-vectors are respectively, $v_\beta = v^\alpha g_{\alpha\beta}$, and, $\dot{v}_\beta = \dot{v}^\alpha g_{\alpha\beta}$. While expressing the instant variation, $\delta x^\alpha \Leftrightarrow \{\delta x_p^\alpha\}$ in the presence of nonlocal correlation, anyone among the infinitesimally differed paths $\left\{x_p^\alpha(\tau)\right\}$ can be selected as reference in the variation expressions. One finds that the nonlocally correlated optimum displacement criterion (1) could be described irrespective of the arbitrarily selected isochronous boundaries (at $\tau_1$ and $\tau_2$) of the evolution course, when integrated part in the variation representation in eq.(1) is a null, and it amounts to the required equality, $(v_\beta \delta x^\beta)\big|_{\tau_2} = (v_\beta \delta x^\beta)\big|_{\tau_1} = C$ (say), for all $\tau_1$ and $\tau_2$ including the limit of $\tau_1 \to \tau_2$. In other words, parameter $C$ in the nonlocally correlated evolution of path family $\left\{x_p^\alpha(\tau)\right\}$ ought to be the same irrespective of the variations and variation boundaries. This is possible only when the nonlocally correlated delocalized evolution course endorses null value of the scalar $(v_\beta \delta x^\beta)$ at all instantly referred space-like surfaces (that is, within instant evolution 3-volume) that are concerned within the evolution course, so that the optimum displacement criterion is endorsed for all time like intervals. The 4-orthogonal connection, $(v_\beta \delta x^\beta) = 0$, corroborating to the optimum displacement criterion can be rewritten by replacing the set of variations $\delta x^\alpha \Leftrightarrow \left\{\delta x_p^\alpha\right\}$ with the corresponding set of space-like unit 4-vectors, $i^\alpha \Leftrightarrow \left\{i_p^\alpha\right\}$ with $i^\alpha \equiv \sum_p c_p i_p^\alpha$, $i^\alpha i_\alpha \equiv -\sum_p c_p^2 = -1$, so that the 4-orthogonal connection takes the form of $(v_\beta i^\beta) = 0$. One can express the space-like unit 4-vector components as $i^\beta \equiv (\delta x^\beta / \delta\tau')$, where $\delta\tau' = \sqrt{-[(\delta x^0)^2 - \sum_j (\delta x^j)^2]}$, with $[(\delta x^0)^2 - \sum_j (\delta x^j)^2] < 0$. The 4-orthogonal connection with displacement 4-vector, $(v_\beta i^\beta) = 0$, represents nonlocal mediation property of the coherently governed geodesics of 4-space. In accomplishing the displacement course instant correlation of the local 4-velocities $v^\mu$ within evolution volume is accomplished by the mediating space-like feature, $i^\beta$. The nonlocal feature, $i^\beta \equiv [\mathrm{i}^0, \ \bar{\mathrm{i}}]$ can be equivalently represented by involving the 3-velocity like components, $\bar{\mathrm{w}} / c \equiv \bar{\mathrm{i}}/\mathrm{i}^0$, as $w^\beta = \ell[1, \bar{\mathrm{w}}/c]$, ($\ell$ being a parameter), so that the 4-orthogonal characteristics, $(v_\beta i^\beta) = 0$ can be rewritten as $(v_\beta w^\beta) = 0$, or, as $\bar{\mathrm{v}} \bullet \bar{\mathrm{w}} = c^2$ with $0 \le \mathrm{v} \le c$, and $c \le \mathrm{w} \le \infty$. The resultant correlation endorses duality of local velocity $\bar{\mathrm{v}}$ and virtually mediating 3-velocity $\bar{\mathrm{w}}$ across evolution volume at any instant of the coherently evolved geodesic displacement course in quantum space. Infinite mediation speed in the virtual correlation of displacement related property accomplishes coherency within instant evolution volume. Concerned evolution harmonics are having instant phase communication property. The lowest value of mediation velocity, $\bar{\mathrm{w}}$ corresponds to the relativistic speed limit ( $c$ ) which is known to be governed by the vacuum properties (dielectric permittivity and magnetic susceptibility). The virtual mediation reveals itself with the relativistic speed limit in radiation relaxation in decoherence that defines the criticality of changeover from quantum to classically governed geodesics in the presence of a large perturbation. In critically perturbed state, the coherently evolved geodesics encounter disentanglement with loss of the nonlocal correlation. The disentanglement will be discussed later in the text.

Now, the nonlocally correlated geodesics meeting optimum displacement criterion in between a pair of arbitrarily selected variational boundaries as expressed in eq.(1) can be rewritten as $\int_{\tau_1}^{\tau_2} \delta\tau' i^\beta [\dot{v}_\beta - v^\alpha v^\sigma \partial_\beta g_{\alpha\sigma}/2] d\tau = 0$, wherein, variation $\delta x^\beta$ is replaced by $i^\beta \delta\tau'$. This result being valid irrespective of arbitrariness in variationally defined parameter, $\delta\tau' = \sqrt{-[(\delta x^0)^2 - \sum(\delta x^i)^2]}$ and also irrespective of arbitrarily selected variation boundaries, $(\tau_1,\ \tau_2)$, it implies that the nonlocally entangled geodesics family evolves obeying the following 4-orthogonality:

$i^\beta[\dot{v}_\beta - v^\alpha v^\sigma \partial_\beta g_{\alpha\sigma}/2] = 0$, or, (for brevity), $i^\sigma (Dv_\sigma / D\tau) = w^\sigma (Dv_\sigma / D\tau) = 0$ (1a).

In (1a), the 4-acceleration associated with displacement in curved space is symbolically expressed as $Dv_\beta / D\tau = dv_\beta / d\tau - v^\alpha v^\sigma \Gamma_{\beta\alpha\sigma}$; $\Gamma_{\beta\alpha\sigma} \equiv \partial_\beta g_{\alpha\sigma}/2$ is space curvature related quantity to be briefly referred as space affinity. $D/D\tau$ takes the simple form, $d/d\tau$ in the special case of flat space for which the metric derivatives, $\partial_\beta g_{\alpha\sigma}$ are all null. The mediation characteristics, $i^\beta v_\beta = 0$, (alternatively, $w^\beta v_\beta = 0$) which complements eq.(1a) leads to 4-orthogonality, $i^\beta \dot{v}_\beta = 0$ because of displacement independency of the nonlocal feature $i^\beta$. Thus, eq(1a) is rewritten in the following form:

$$i^\beta A_\beta = w^\beta A_\beta = 0 \qquad (1b),$$

where, $A_\beta = (v^\alpha v^\sigma \partial_\beta g_{\alpha\sigma}/2) \equiv v^\alpha v^\sigma \Gamma_{\beta\alpha\sigma}$. Eq.(1b) describes nonlocal constraint of the affinity field $\Gamma_{\beta\alpha\sigma}$ of virtually mediated quantum space. In the nonlocally entangled state, as shown in eq(1b), the displacement 4-vectors of nonlocally correlated geodesics, namely, $v^\alpha = \sum_p c_p v_p^\alpha$ conform to the constant length, $v^\alpha v_\alpha = \sum_p c_p^2 = 1$. The described entanglement of space affinity with nonlocal mediation velocity in eq.(1b) of quantum space, however, breaks down when this state is externally perturbed. With the inception of local perturbation in quantum space, the affected part of space can be deviating from coherent description (1a) as briefly discussed in the following paragraph.

Coherently evolved space once perturbed locally, the nonlocal correlation need not disappear there immediately with the inception of external perturbation. This is because of prompt mitigation by ubiquitous defense from the large unperturbed part of field. The mitigating defense incessantly reacts on the local perturbation to help sustain nonlocally mediated instant correlation until the coherence in evolution harmonics is critically perturbed with relaxation loss. The defense reaction on external perturbation reveals itself as recoil from the momentum loss in power relaxation at the decoherence criticality and this aspect will be brought out in details for electrodynamics as a special case (refer Part-B). Driven by increasingly high perturbation the ubiquitous defense safeguard to evolution harmonics can sustain mitigation effectively over a brief onset course, until the evolution harmonics suffering from increasingly high fluctuations in perturbation are ultimately deprived of access of the nonlocal commands in thoroughly beaten state to result in decoherence. The brief onset to relaxation event will be reflected in equation (1) by abrupt increase in randomness of the instantly governed nonlocal correlation parameters, $\{c_p\}$, which at decoherence criticality culminates with complete arbitrariness in the variation $\delta x^{\alpha} = \sum_p c_p \delta x_p^{\alpha}$. Once the perturbation is locally impressed upon there is ever increase in randomness in the correlation parameters until the randomness thwarts the nonlocal influence. The set of instant correlations, $\{c_p\}$, that are commonly referred by displacement parameter, $\tau$, are now additionally governed by stochastic factors responsible for the random fluctuations. Under sustenance of ubiquitously mitigating defense during the brief onset period, the mediation course will be revised by the null value of stochastic average product, $\langle v_{\mu}\delta x^{\mu}\rangle$ at an instant, rather than instantly referred null product $(v_{\mu}\delta x^{\mu})$ irrespective of the paths, $p$. Null product $\langle v_{\mu}\delta x^{\mu}\rangle$ will help eliminating the integrated term present in the averaged representation of eq.(1), which is used to describe the brief onset course to perturbation criticality. Thus, irrespective of the arbitrary variation $\langle \delta x^{\mu}\rangle$ the virtually mitigated dynamic evolution can be represented there in the averaged form as

$$\langle \dot{v}_{\beta} - v^{\alpha} v^{\sigma} \partial_{\beta} g_{\alpha\sigma} / 2 \rangle = 0 \qquad \text{(1c).}$$

Eq.(1c) involving all local terms represents the classically described geodesics of space.

A1.1 Exotic quantum vacuum versus cosmic background vacuum

Mediation property, $w^{\mu}$ in its spectral representation as $w^{\mu} \equiv \sum_k c_k w_k^{\mu}$, $w_k^{\beta} = \ell_{\mathrm{k}}[1, \bar{\mathrm{w}}_{\mathrm{k}}/c]$, ($\bar{\mathrm{w}}_{\mathrm{k}}$ being spectral velocity, $\omega\bar{\mathrm{n}}_{\mathrm{k}}/\mathrm{k}$, with $\ell_{\mathrm{k}}$, as a parameter) shows its 4-orthogonal connection with the spectral 4-coordinates $k^{\alpha} \equiv [\omega/c, \bar{\mathrm{k}}]$; $\bar{\mathrm{n}}_{\mathrm{k}}$ being unit vector along $\bar{\mathrm{k}}$, ($\mathrm{n}_{\mathrm{k}}^2 = 1$). With the perennial 4-orthogonal connection, $k_{\beta} w_k^{\beta} = 0$, it may be noted that the coherent evolution property of geodesics shown in (1b) continues to remain valid under the harmonic transformation of space metric as $g'_{\alpha\sigma}(x) = [1 + O(x)] g_{\alpha\sigma}(x)$, with $O(x) = \sum_k c_k \exp(-ik_{\mu}x^{\mu})$ + c.c. Transformed product $A'_{\beta} w^{\beta}$ of eq.(1b), that is, $A'_{\beta} w^{\beta} = w^{\beta}(v^{\alpha} v^{\sigma} \partial_{\beta} g'_{\alpha\sigma}/2)$ results additional term, $g_{\alpha\sigma}(w^{\beta}\partial_{\beta} O)/2$ which will be a null due to the perennial 4-orthogonallity, $k_\beta w_k^\beta = 0$ in the expression, $w^\beta \partial_\beta O = \sum_k c_k^2 w_k^\beta (-ik_\beta)\exp(-ik_\mu x^\mu) + \text{c.c.}$. The equality, $w^\beta \partial_\beta O = 0$, when compared with the mediation property, namely, $v^\alpha w_\alpha = \sum_k c_k^2 v_k^\alpha w_{k,\alpha} = 0$, one gets the following proportional connection:

$$v_\beta \propto \sum_k c_k (-ik_\beta)[\exp(-ik_\alpha x^\alpha)] + \text{c.c.}$$

According to the proportional connection, the coherently evolved geodesics in quantum space possess harmonic displacement property. Displacement, $v^\alpha = \sum_k c_k v_p^\alpha$ conforming to the constant length, $v^\alpha v_\alpha = \sum_k c_k^2 = 1$, the above proportional connection can be rewritten in scalar form as $\sum_k |c_k|^2 \propto \sum_k |c_k|^2 (k^\alpha k_\alpha)$, that is, as $\sum_k |c_k|^2 [k^\beta k_\beta - (2\pi/\lambdabar)^2] = 0$. The constant quantity, $(2\pi/\lambdabar)$ having dimensionality of inverse length arises out of the proportional connection. The wave dispersion property of nonlocally evolved geodesics refer length element, $\lambdabar$ as the basic feature of quantum space concerned. The dispersion characteristics endorse freely evolved harmonics. Arrived characteristics of coherently evolved geodesics field entails virtual exchange of quantized energies ($\hbar\omega$) of the evolved harmonic modes as given by $\hbar\omega = \sqrt{c^2\hbar^2(\mathrm{k}^2 + 4\pi^2/\lambdabar^2)} \geq 2\pi c\hbar/\lambdabar$, ($0 \leq |\mathrm{k}| \leq \infty$), where, $2\pi c\hbar/\lambdabar \equiv E_{vac}^0$ (say) is the lowest mode energy. The modes possess finite energies with a nonzero value of the coherent dimension, $\lambdabar$. The wave dispersion property thus leads to the fact that the space, which is solely described by the geodesic field, is composed of a large numbers of freely evolving harmonic oscillators modes having frequencies at and above the lowermost limit, that is, $\omega \geq 2\pi c/\lambdabar$. In the absence of any other fields, the coherent description of quantum space applies to ideal vacuum state, the quantum vacuum (QV). QV is distinct from the ideally extrapolated state of cosmic space as its recessional property is taken in the limit of null density of gravitating matters. Within the state of nothingness, QV is displayed with the dimensional identity, $\lambdabar$, and energies $\hbar\omega$ as expressed above; $E_{vac}^0$ being the lowermost limit.

A quantum mass of $2\pi\hbar/c\lambdabar$ corresponding to lowest mode energy of the vacuum field's evolution harmonics cannot be too large to camouflage the dimensional identity within event horizon of radius, $2\pi G\hbar/c^3\lambdabar$. This implies the minimum value of dimension ($\lambdabar$) as $\lambdabar_{\min} = \sqrt{4\pi\hbar G/c^3}$; $\lambdabar_{\min}$ being about 3.5 times Planck's length (~1.616x10$^{-35}$m; $G$ being the gravitational constant). This limit suggests that quantum space having energy equivalent mass density in excess of $2\pi\hbar/c\lambdabar_{\min}^4$ (~2x10$^{95}$kgm$^{-3}$) will remain camouflaged as invisible gravitating source which has relevance within event horizon of a black hole. On the presumption that quantum spaces are describable with a basic dimensional element having relevance in the description of time evolved three dimensional world of observables, the corresponding $\lambda$ value is expected to be higher than $\lambda_{\min}$. It is prudent to have $\lambda$ which is relevant in the observed kinematic properties of space. $\lambda$-value can be arrived through a comprehensive analysis of recessional kinematics associated with the background field of cosmic space. For the analysis one thus takes into the consideration that cosmic background vacuum, unlike the exotic quantum vacuum, resonantly/quasi-resonantly supports the physical world of all types of evolution harmonics covering the continuous frequency range, $0 \le \omega \le 2\pi c/\lambda$, where the uppermost limit is expressed by using the basic length dimension of the observable three dimensional gravitating space in recession. As indicated already, the observable space apparently posing as a virtual 3D lattice-like quantized structure supports the physical world in entirety. As basic cell element of the virtual lattice, this analysis with the help of classical field equation of general relativity evaluates the quantized space volume, $\lambda^3$ in ideally extrapolated state of time evolved virtual lattice-like space towards the limit of null density of the gravitating matters. $\lambda$ value evaluated thereof as 7.266x$10^{-35}$m is about five times Planck length. This aspect will be elaborated later in supplement, S-1.

Quantum vacuum field with its lowest energy state of $2\pi c\hbar/\lambda \equiv E^0_{vac}$~2.5x$10^9$J (as per evaluated $\lambda$~7.266x$10^{35}$m), is constituted of enormously high energy harmonic modes as compared to the resonance energies of all known elementary particles. Vacuum energy density in the time evolved 3-D space is of the order of $E^0_{vac}/\lambda^3$~$10^{112}$J$m^{-3}$ which corresponds to repulsive field pressure (N$m^{-2}$) of that order. Ubiquitous mighty quantum vacuum (or, false vacuum) is thus exotic to the observable physical world which is existing stably/quasi-stably within the background vacuum of cosmic space. The mighty QV field being composed of ultrahigh energy harmonic oscillators, the stable/quasi-stable existences of worldly objects are ruled out under its direct influence.

A1.2 Understanding recessional aspects of cosmic world

In view of the contrasting features of the two types of vacuum fields, namely, the exotic QV and the background vacuum (BV) of cosmic space, it is necessary to understand their coexistence and mutual interaction in order to comprehend the ubiquitous recessional characteristics of the gravitating space. Apparently, the virtual lattice-like 3D network of the observable space supporting the gravitating cosmic objects is undergoing time evolved recession with definable kinematics for over billions of years evolutionary past following the symbolic big banged beginning presumably from ubiquitously fluctuating field of mighty quantum vacuum within nothingness. The expanding lattice ordered space network has been cooling ever since the symbolic big-bang beginning. From ultra-hot past history of space, the currently exhibited thermal state of 2.725±0.00013 K corresponds to the nearly uniform cosmic microwave background field. The expansion of lattice-like gravitating space with thermal cooling is indicative of the expense of internal energy in volume displacement against inward pull of gravity. Presumably, the spontaneity in expansion of lattice ordered state within precursor QV field is because of ubiquitous mitigation property of the mighty field to weed out the local heterogeneity from the gravitationally perturbed lattice that came into being in big-bang. In mitigation, whatever field energy is used in driving the lattice expansion is fully recovered as compressive work on the repulsive field itself. Energy density and gravity field strength of the lattice-ordered space fall progressively in the field mitigated expansion. Mighty vacuum field having ultrahigh repulsive energy density, however, cannot directly mitigate on the gravitationally perturbed background vacuum field because of high kinetic impediment due to strikingly different quantum order of the perturbed virtual-lattice being mitigated. The two types of vacuum fields with their distinguishable evolution harmonics can interact across inescapable energy barrier thus inherited. Mighty QV field mitigates on the gravitationally perturbed lattice-like space indirectly by quantum tunneling of field energy. Kinematic impediment in the tunneling course largely cuts down the ultrahigh repulsive influence of QV field on the gravitationally active space and also cuts down the inherently present field fluctuations of high orders.

In the mitigation across energy barrier the tunneled field quanta result in manifestations of lattice-cells and their immediate integrations within the barrier isolated 3D-network of virtual lattice-like gravitationally active space uniformly throughout, enlarging thereby the lattice with consequential increase of inter-separations of the homogeneously distributed gravitating matters in cosmic scale. In the kinematic expansion course of lattice, each field quantum involved in mitigation thus leads to the volume displacement work against internal gravity field together with stress accumulation within mitigated space. In mitigation, for each newly manifested and integrated cell of quantum volume $\lambda\!\!\!^{-}{}^3$, the field spares a quantum of energy $E^0_{\mathrm{vac}}$ and recovers that energy as displacement work associated with the expansion by $\lambda\!\!\!^{-}{}^3$ volume of the barrier isolated lattice network against repulsive external pressure, $p^0_{vac}$ of mitigating field; displacement work being, $\rho^0_{vac}\lambda\!\!\!^{-}{}^3 = E^0_{\mathrm{vac}}$, ($\rho^0_{vac} = -p^0_{vac}$). In keeping with the requirement of energy recovery in mitigation, QV field does not involve more than one quantum of energy per manifested cell volume. The volume dilation is again associated with stress accumulation in the time evolved gravitating space with progressively increased internal pressure within the barrier isolation. Involved volume displacement against gravity field and stress accumulation together is accomplished by the reduction of internal energy with consequential thermal cooling. In the long course of sustained lattice dilation, the mighty QV field by the ubiquitous mitigation will finally weed out the lattice ordered state of the idealized background vacuum field while approaching towards null density of matters where the gravity field becomes too weak to sustain the lattice ordered state of cosmic space with internal pressure approaching the limit, $p_{vac}^{0}$. The kinematic barrier existing between the two fields because of their strikingly different quantum orders becomes insignificant to the repulsive pressure. Long held meta-stability in the form of recessional space of gravitating matters derived out of field fluctuation catalyzed local event within mighty precursor field would be nonexistent in the long run.

The billions of years long recessional history of cosmic space thus appears to be due to the sustenance of lattice-like network in a state of meta-stability acquired by gravitation fields of the accumulated stresses as well as point-like defects due to the subatomic species in their locally ordered atomic/ionic states distributed within virtual lattice. The stresses and point-like defects together as gravitating matters prove their presences by the resonant/quasi-resonant evolution harmonics within the quantum space. (Evolution harmonics of the gravity bound stresses are, however, matters of future investigations). In the long evolution course, the locally ordered subatomic species got segregated by their gravitational influences leading to formations small dust particles, and subsequently to large segregates in the different forms of cosmic objects, such as planets, stars, galaxies and clusters of galaxies, and super-heavy masses. Initially delocalized stresses tend to concentrate locally around high density of the visible matters. Accumulated stresses get segregated according to local gravity fields of the heavenly bodies. According to the local densities of matters/stresses the gravitational field strength is reflected by distortions of the geodesic network. With the ultrafast response time of $\lambda\!\!\!^{-} / c$ (~$10^{-43}$s) the virtual lattice ordered network of space non-interferingly supports gravitating objects of all the forms having orders of magnitudes slower evolution harmonics.

A1.3 <u>Mighty QV field as precursor of gravitating cosmic space</u>

Two aspects are apparent in the above discussion on space recession kinematics of cosmic world by ubiquitous QV field mitigation:

(a) Apparently, the smallest space length envisaged by Max Planck shows relevance in the descriptions of mitigating QV field and of mitigated background field of time evolved 3-D cosmic space (real vacuum). Both the fields are made of the elemental length ($\lambda\!\!\!^{-}$) with subtleties in quantum orders in space evolutions for the respective cases. As shown in supplement, S-1, consideration of the recession kinematics of cosmic space leads to evaluation of the relevant length dimension.

(b) It is quite possible that the gravitating cosmic space constituted of virtual lattice ordered harmonics ($0 \le \hbar\omega \le 2\pi c\hbar / \lambda\!\!\!^{-}$) was derived out from high energy states of the mitigating QV field ($\hbar\omega \ge E_{\mathrm{vac}}^{0}$) in the distant past. However, as will be addressed now, in making comprehension about such beginning of the gravitating cosmos hurdles are no less. For qualifying the beginning from QV field as precursor, which is composed of harmonic oscillators of ultrahigh energies, one encounters difficulty in understanding the changeover of the mighty oscillators' field to that of the virtual lattice ordered quantum space supporting gravitating matters.

With an exceptionally low but of finite probability, mighty precursor in locally high excitations in inherent field fluctuation might have taken altered de-excitation course through evanescent ordering of excited modes while releasing out the borrowed excitation energy promptly in favor of attaining comparatively more stable states under harmonic potential of embryonic lattice-like ordered space which is gravitationally active under the barrier isolation from surrounding precursor. QV modes in high excitations only can avail the said alternative de-excitation course in realizing the lattice ordered space with strikingly different evolution harmonics locally under the barrier isolation. High barrier energy isolation ($E^*$) impedes fast decay of the virtual lattice order space back to the unperturbed QV state and thus give a chance of other competing kinematics in taking over to altered course of attaining stability under the harmonic potential. As will be made clear in the text, this to happen requires an ultrahigh field fluctuation so that a significant number of momentarily excited field modes can participate in taking the altered route of their initial ordering as a quantum lattice in evanescence under the barrier isolation. As will be shown subsequently, highly stressed embryonic space once formed in fluctuation catalyzed event immediately undergoes QV field mitigated intricate relaxation course involving the manifestation of gravitating matters through spontaneous symmetry breaking event of the expanding 3D virtual-lattice network in ultrafast space inflationary mode with super-cooling. Indicated symmetric breaking course of the evanescently ordered space closely follows that of the false vacuum state as reported in the literatures [11]. As discussed in supplement, S-3, highly stressed lattice ordered quantum space inherently supporting transiently evolved scalar fields has the potential to relax out the stress energy under suitable boundary condition through breaking of the fully symmetric state through generation of field resonance of gravitating masses like Higgs boson. Manifestations of fermionic particles then follow through well known Higgs mechanism.

In fluctuation catalyzed events within QV field, no soon a group of $N'$ numbers of excited harmonic modes of energy, $E^0_{vac} + \Delta E$, ($\Delta E \geq E^*$) have made transient access across barrier ($E^*$) to the virtual lattice ordered states of embryonic space, a subgroup $N''$ among them can undergo fast de-excitation to lower energy states $E'$ ($0 \leq E' < E^0_{vac}$) of the virtual lattice in paying back the entire borrowed energy, $N'\Delta E$. The energy pay back step occurs within the time period of $\hbar / [(E^0_{vac} + \Delta E) - E']$, which is well within the time span of the short fluctuation course ($\sim \hbar / \Delta E$). The energy pay off step follows the conservation equation as, $N'\Delta E = N''[(E^0_{vac} + \Delta E) - E']$ and this leads to, $N' / N'' = [1 + (E^0_{vac} - E') / \Delta E] > 1$, where the ratio, $N' / N''$ never exceeds the quantity, $(1+E^0_{vac}/E^*)$. In parallel to the energy pay off step, the remaining $(N'-N'')$ number of excited modes also de-excite into the virtual lattice ordered states in redistributing their energies among all the phonon modes with maximum accession of internal states ($0 \le E' < E^0_{vac}$) attaining averaged energy of $E^0_{vac}$ per mode ($E^0_{vac} = 2\pi c\hbar/\lambdabar$).The lattice-like embryonic space thus realized locally within the fluctuation perturbed precursor with the averaged energy $E^0_{vac}$ corresponds to thermally elevated state at $E^0_{vac}/k_B = T_0$ ~2x10[32] K, ($k_B$ being Boltzmann constant). This barrier isolated embryonic space of high meta-stability as reflected by the averaged energy $E^0_{vac}$ of phonon mode, is analogous to the generally referred false vacuum as intermediate, which promptly takes the field mitigated ultrafast stress relaxation course involving symmetry breaking events with the manifestations of gravitating matters under abrupt lowering of phonon energies in inflationary space expansion as part of the course. At the onset of relaxation, the highly stressed lattice ordered embryonic space generates gravity field according to the averaged phonon energy, the magnitude of which at spherical surface of kernel is of the order of $F_G = -GE_T/c^2R_k^2$, ($R_k$, being kernel radius) wherein, energy $E_T$ corresponds to the total gravitating sources within the kernel; matters, stresses, and radiation taken together will corroborate approximately to $E_T \sim N'k_BT_0$.

As mentioned already, the mitigation rate of QV field through barrier tunneling of field quanta followed by the manifestations and integrations of the mitigated cells within the virtual lattice-like 3-D network of space is expected to be slow as compared to backward kinematics of kernel's resolution back to the precursor field. Nevertheless, the whole kinematic course of field mitigated enlargement of the virtual lattice ordered space can be faster when the tunneling rate of QV field quanta is facilitated under effectively suppressed barrier as can be achieved by augmenting the kernel's gravity field ($\bar{F}_G$) so as to compensate out the barrier energy ($E^*$) effectively by the displacement work involved in the transport of field energy across barrier width ($b$) under field, $\bar{F}_G$. For a critically large dimension of the ultra-hot kernel, gravity field will be high enough for the required compensation. Critical dimension ($R_k = R_{cr}$) of the barrier isolated gravitationally active embryonic kernel that meets barrier compensation criterion can be arrived from this consideration at the extremely high temperature as $E^* \sim b\left|\bar{F}_G\right|(E^0_{vac}/c^2)$. Now, as shown in supplement, S-2, a quantum of field energy $E^0_{vac}$ once penetrates the barrier it will lead to the mitigated settlement of a new cell within the lattice-like space network meeting thermodynamic spontaneity when the ultra-hot embryonic space possesses at least a radius of about $1\text{x}10^{-33}$ m ($14\lambdabar$) accommodating $1\text{x}10^4$ lattice cells. With this radius the internal field $\bar{F}_G$ will meet the barrier compensation criterion for a minimum width of $b \sim \lambdabar$ when $E^*$ corresponds to $\lambda\left|F_G\right|(E_{vac}^0/c^2)$ which is of the order of $9E_{vac}^0$. The evanescently formed embryonic lattice ordered ultra-hot kernel developed with the critical radius of about 1x10$^{-33}$ m is normally expected to invite the coherent field mitigated space inflationary relaxation course. However, in the events of fluctuation catalyzed manifestation of the critically formed kernel within QV field there are other things to be considered in this regard. The fluctuation catalyzed course produces a large crowd of kernel spaces in the form of an ultra-hot froth in the fluctuation affected locality. Once a kernel emerges out of a high fluctuation, ubiquitous mitigation by the mighty quantum vacuum field promptly reacts to weed out the locally appeared heterogeneity and in that process it invites more fluctuation events of ultrahigh orders in that locality leading to manifestation of a large crowd of equally ultra-hot subcritical kernels along with one or a few critically developed ones in the form of a locally formed forth. Relatively high gravity field ($\bar{F}_G$) of a critically developed kernel ($\geq 10^{-33}$m) helps thinning down the subcritical ones of immediate vicinity by attracting them through field mitigated fast diffusion having time constant of the order of $d_k^2(\hbar/2m_k)^{-1}$, where $m_k$ and $d_k$ are respectively the mass and diameter of diffusing smaller kernels of the crowd. In the diffusion aided coalescence process, the critically developed one of radius, $R_{cr}$ (say) is thereby getting oversized to $mR_{cr}$, ($m>>1$) within the characteristic time period. No soon the surrounding crowd thins down through the coalescence, which is estimated to be within 1.5x10$^{-37}$s to 5x10$^{-35}$s, the coherent field mediated ultrafast space inflationary mitigation starts in the barrier suppressed state of the oversized kernel having average temperature $T_0$ ~ 2x10$^{32}$K. The ultrafast course is expected to continue till the internal gravity field of the inflated and super-cooled kernel space becomes too weak to suppress the barrier effect.

In the space inflationary course, the uniformly mediated field mitigation throughout the thermally as well as gravitationally perturbed lattice-like 3-D space of radius $mR_{cr}$ leads to ultrafast settlements of mitigated cells which are largely having no thermal history. Settlements occur effectively at all the three interfaces for each cell of the cubic lattice ordered space. Thus, within a quantum period of $\lambda/c$ ~10$^{-43}$s, a lattice of $n$-cells multiply effectively by $3n$ and therefore, after $N$ quantum periods in succession the coherent field mitigation leads to enlarged lattice of $n(3^N-1)/2$ cells and thereto the super-cooling of space. With the multiplication factor of $M=(3^N-1)/2$, thermally equilibrated state of the super-cooled space would endorse a temperature of $T_0/M$. As indicated already, no soon the (boson-like) phonon field of ultrahigh energy starts passing through the abrupt super-cooling course involving steep temperature gradient across the ultra-hot lattice-like space, the thermally stressed state encounter metamorphic transformations involving the symmetry breaking events in successive stages of the space inflationary course leading to the manifestations of nucleonic species in dense (opaque) plasma state by the time temperature reaches to around $10^{12}$K. That the inflated space should support existence of the lightest particle like electron anti-neutrino among nucleonic matters implies that the spherical space has attained a dimension of the order of resonance length for the lightest specie which is of the order of about 1x$10^{-6}$m (mass less than 1eV [12]). The correspondingly attained kernel space volume can accommodate about $10^{84}$ numbers of closely packed lattice cells. This suggests that the embryonic kernel of radius $10^{-33}$m ($R_{cr}$) and temperature ($T_0$) of about 2x$10^{32}$K barely accommodating $10^4$ lattice-cells (ref supplement, S-2) has undergone overall cell multiplication of $10^{80}$ within the brief period of about $10^{-34}$s during which there is inflationary space expansion and supercooling from the interim overgrowth dimension (radius $mR_{cr}$). Now, consideration that the super-cooled temperature is $10^{12}$K, then by using $M = (3^N - 1)/2$ in the equality $T_0/M$ =$10^{12}$K one obtains the quantum steps of $N$~41.9, and accordingly, the inflationary factor of $M$~2x$10^{20}$ occurs within the brief period of $41.9(\lambda/c)$ seconds. This suggests that a significantly large numbers of sub-critically manifested ultra-hot kernels in the fluctuation affected locality have coalesced forming the oversized kernel (radius ~$10^{-13}$m) before striking onset for the inflationary space expansion with the super-cooling.

During the space inflationary super-cooling course of the highly stressed lattice, the subatomic species as relaxation products in plasma state are manifested in the symmetry breaking events which presumably involve in-situ interactions of Higgs resonance with fermionic fields in acquiring masses of fermions. For yet unclear reason, field mitigation in the ultrafast cooling step compromises for a biased course of slightly unequal manifestations of matters and anti-matters. Presumably, in the presence of matter-antimatter recombination heats the space inflationary super-cooling kinematics in the biased course of the symmetry breaking steps could optimally conform to the observable yield of visible matters as about 5%. As discussed in supplement, S-3, under the influence of bosonic and fermionic fields the pseudo space-curvature definable with asymmetric stress can lead to biasness in the manifestations of fermions and their anti-particles in the symmetry breaking events of field mitigated ultrafast relaxation course of the barrier confined lattice ordered space derived out of fluctuation catalyzed QV field.

Inflations to enormously enlarged super-cooled space abruptly weakens the gravity field ($\bar{F}_G$) so much within a short while (<$10^{-32}$s) that the barrier suppression criterion is difficult to be met in the subsequence, particularly in the wake of reheating from exothermic energy release in the metamorphic transformations. The in-situ thermal surge results in abrupt deviation from the coherently mediated space inflationary course by a sharp drop in space recession as well as cooling rate. In the post inflation period, space containing the metamorphic transformation products in the form of ultra-hot dark plasma (temperature $\leq 10^{12} K$) of nucleonic matters undergoes barrier regulated decelerated space recession kinematics under continued mitigation of QV field across progressively increasing effective barrier height from the critically compensated state. With the revival of effective barrier height, the internal field of the gravitationally perturbed space falls off progressively according to the recession kinematics. The recession involves volume displacement work in making rooms for the mitigated settlements of new cells along with accumulation of stress within virtual lattice ordered gravitating space. Until the barrier could revive appreciably in making notable impediment in field mitigation the space recession rate remains remarkably high. Accordingly, the very initial part of post inflation course is covered with fairly large values of Hubble parameter $H(t)$ under a moderated deceleration in space recession before attaining the nearly uniform decelerated space recession which then stretches over the several billions of years. The large initial $H(t)$ of the post inflationary period has bearing on the directly measured recession kinematics of astronomical objects (such as type Ia supernovæ) as noted in the recent time [10]. Recently observed recession characteristics of supernovæ in distant galaxies do reveal higher magnitude of Hubble's constant (about 73±1 km $Mpc^{-1}s^{-1}$) as compared to the theoretically predicted one (about 67.5±0.5 km $Mpc^{-1}s^{-1}$) which is based on the use of 'Lamda-CDM' model on the noted CMB data [13]. Subsequent to the initially radiation dominated cosmic evolution, the billions of year long matter dominated decelerated recession of the gravitating space involves barrier impeded mitigation rate by the repulsive effect in mitigation by precursor field. With the sustained weakening of internal gravity field in space recession the virtual lattice ordered space network gradually weakens and this leads to a stage when repulsive effect in field mitigation starts showing up with progressively faster space recession rate. This is evidenced in the recently noted trend of speediness in the rate [10]. In the long run the repulsive influence will outwit barrier impedance as well as the existence of virtual lattice ordered space to result in transformation of the ordered state back to the state of mighty quantum vacuum.

A1.4 <u>Relative proportions of invisible gravitating matter and repulsive energy</u>

In the cosmic space recession, QV field mitigated incorporation of quantum cells within the virtual lattice ordered state involves volume displacement work against internal gravity field. Besides there is stress accumulation at a finite growth rate of internal pressure. The repulsive influence of mitigating field is registered in the rate of volume displacement against the gravitational compression pressure, $p(t)$. For the nonlocally governed recession occurring uniformly throughout the space, one expresses the kinematics by taking note of Newtonian action and reaction involved in mitigated incorporation quantum cells within the lattice that expands accordingly. In making rooms for the cell incorporation within 3-D lattice-like virtual network, the QV field mitigated pressure thrust on the equally shared interfaces of each lattice-cell manifests as kinematic acceleration with which the interfacial gaps widens in making rooms for cell incorporations. The concerned displacement kinematics on a lattice-cell is thus represented by the proportionality: $[\rho' a^3 / c^2]\ddot{a} \propto pa^2$, where, $\rho'(t)$ is average energy density of the virtual lattice ordered space network, and $a(t)$ is the time evolved scale factor in space recession. Thus, one writes $p(t) = K\rho' a\ddot{a}$, ($K$ being the proportionality constant). Accordingly, internal pressure, $p(t)$ evolves in the course of field mitigated space evolution as $\dot{p} = Ka^2\rho'(\dot{a}/a)(\ddot{a}/a) + Ka[d(\rho'\ddot{a})/dt]$ and this expression to its first approximation can be written as $\dot{p} \approx K\rho' a^2[(\dot{a}/a)(\ddot{a}/a)]$, where the contribution of kinematic jerk in the time evolved pressure appeared in the second term of right hand side is dropped as it contributes insignificantly compared to the acceleration term. (Jerk's presence though negligible in comparison to the kinematic acceleration, it has facilitated segregation of the weakly interacting dust matters towards the formations of massive heavenly bodies). Noting that time evolved scale factor, $a(t)$ is endorsed by the recession kinematics of heavenly bodies uniformly within the observable cosmic sphere of radius, $R$ one writes the correspondence in the scaling as $(\dot{a}/a) = (\dot{R}/R) \equiv H$, where $H(t)$ is the Hubble's parameter in cosmic space evolution course. In terms of $H$-parameter, the kinematic acceleration can be written as $\ddot{a}/a = H^2 + \dot{H}$, which with the use of $\dot{H} = -H^2(1+q)$, takes the form of $\ddot{a}/a = -qH^2$. For the decelerated recession history, ($\ddot{a}/a < 0$), as one represents the acceleration parameter of space as $q = -(\ddot{a}/a)/(\dot{a}/a)^2$, $q$ behaves as a positive quantity. Now, by using the equality, $\ddot{a}/a = -qH^2$, one rewrites the earlier obtained pressure expression, $p(t) = K\rho' a\ddot{a}$, as $p(t) = -K\rho' qa^2H^2$. Also, the time derivative of pressure so far expressed as $\dot{p} \approx K\rho' a^2[(\dot{a}/a)(\ddot{a}/a)]$ now takes the form of $\dot{p} \approx -K\rho' qa^2H^3$. Negative $p$ reflects the repulsive influence of mitigating field on internal pressure and negative $\dot{p}$ in the space recession corresponds to increasing repulsive influence of the field with time. Ratio, $p/\dot{p}$ is thus approximately governed by the inverse of recession speed $H$ as $p/\dot{p} \approx 1/H$. The repulsive influence of mitigating QV field is registered within the virtual lattice-like cosmic sphere as joint contributions of two power terms, namely, the power, $p\dot{V}$ concerned with volume displacement against gravity field and power, $V\dot{p}$ which is related to accumulation rate of the gravity bound stresses; $V$ being $4\pi R^3/3$. Cumulative stress, $\int V\dot{p}dt$, corresponds to a time evolved effective mass which is invisible in the gravitating space that contain about 5% visible mass. Now, considering that the two invisible power terms involved in nonlocally mitigated recession of the virtual lattice-like cosmic sphere is endorsed uniformly throughout the space, their ratio is expressible as $[p(\dot{V}/V) \div \dot{p}]$, which on the substitutions of $(\dot{V}/V) = 3\dot{R}/R = 3H$, and $p/\dot{p} \approx 1/H$, can be rewritten as $[p(\dot{V}/V) \div \dot{p}] \approx 3$. Thus, to a first approximation (neglecting the jerk contribution), the above analysis shows that the two invisible power signatures in decelerated space recession follow the constant ratio of $3:1$, which remained valid all through the field mitigation history irrespective of the time evolved recession parameter $H(t)$. It shows that the two mitigation signatures in their cumulative representations for the respective energies over the entire space evolution history corroborate approximately to the constant proportion of $3:1$. If one considers that the attributions of accumulated stresses and stored work within the virtual lattice-like space correspond respectively to gravitating dark matter and repulsive dark energy one finds that they are in the ratio of about 1:3. Since the cosmic beginning from QV as the precursor field, these invisible energy signatures along with the visible matters (roughly 5%) are the essential features of field mitigated cosmic space recession over 13.8 billion years long evolution period. The evolution maintains homogeneous and isotropic distribution of visible matters in cosmic scale and as per the presented analysis the ratios of visible matter **:** dark matter **:** dark energy are of the order of 5**:**24**:**71. The presence of repulsive dark energy prevents gravitational crunch while gravitationally active dark matter with its tendency to concentrate around regions of high gravitational activity moderates on the dispersion tendencies of orderly segregates of visible matters existing in the form of galaxies and galaxy-clusters.

A1.5 Summary

Mighty quantum vacuum field as precursor to cosmic space incessantly mitigates on the gravitationally perturbed state uniformly all through across high kinematic barrier. Within the nonlocal doctrine, mitigation keeps up homogeneous and isotropic nature of the time evolved cosmos meeting causality in all physical events within the virtual lattice ordered quantum space having coherently evolved geodesic network in near criticality of its gravitationally perturbed state. While quantum states of a physical system in cosmic space encounter loss of coherence in the presence of external perturbation the coherently evolved geodesic field remains undeterred unless the perturbation is exceptionally high. The coherently evolved harmonics involved in mediating ubiquitous correlation in a physical system belongs to the family of harmonics $(0 \leq \hbar\omega \leq 2\pi c\hbar / \lambda\!\!\!^{-})$ describing geodesic field of the lattice ordered quantum space. Thus, in a way the geodesic field providing logistic support for the instant nonlocal correlation of a coherent system also helps in maintaining causality in the event of instant collapse of wave vectors of the quantum state in presence of external perturbation. This is true even for quantum mechanically entangled pair of distantly separated photons/particles when subjected to local probing on one of the entangled partners. It is again worth mentioning that in absence of the barrier's isolation, mighty precursor field by its nonlocally governed mitigation would have denied the permanence of lattice ordered cosmic space along with all physical displays. Without the barrier's isolation, sustenance of the gravitationally perturbed lattice-like cosmic space over billions of year long recession history had been impossible. The barrier isolation also does away with the manifestation possibility of new kernel spaces with their inflationary developments within the gravitating space itself, which otherwise is possible within the ultrahigh fluctuation prone mighty field. Barrier's presence helps the existence of observer as well as the observed in the virtual lattice ordered space which has been undergoing recession steadily even with the utterly low density of visible matters ~4x10$^{-28}$kgm$^{-3}$.

**PART-B**

B1.0 <u>Dynamic evolution of point-like charge in electromagnetic field</u>

Nonlocally mediated coherent course of dynamic evolution of interacting field and particle can be described in similar way as has been carried out for the ubiquitously correlated evolution of geodesics of quantum space by using variation optimization of the (virtual) displacement course. Thus, in the case of electrodynamic evolution, the nonlocally correlated optimum displacement course in between arbitrarily selected pair of space like surface-boundaries is expressible as

$$\int_{\tau_1}^{\tau_2}\delta\left(\sum_p c_p^2(g_{\alpha\beta}dx_p^\alpha dx_p^\beta)\right)^{1/2}+\int_{\tau_1}^{\tau_2}\delta\left(\sum_p(c_p g_{\alpha\beta}dx_p^\beta)\int[c_{p'}(q/m_0c^2)A^\alpha(x_{p'})\delta^3(x_{p'}-x_p)d^3x_{p'}]\right)=0 \quad (2).$$

In eq.(2), the second term involving field-particle interaction leads to deviation in the correlated displacement course from that of field free evolution of the particle. The arbitrarily selected pair of space-like variation boundaries are having the time-like separation, $(\tau_2-\tau_1)$. Four vector, $A^\alpha(x_{p'})$ represents 4-potential of electromagnetic field; $q$ and $m_0$ respectively being charge and mass of point-like particle having charge density, $q\delta^3(x_{p'}-x_p)$ in space. Weight factors $\{c_p\}$ associated with the linearly combined displacement components $dx^\alpha \Leftrightarrow \{dx_p^\alpha\}$ as $dx^\alpha=\sum_p c_p dx_p^\alpha$, conform to the instant correlation characteristics, $c_p c_{p'}=\delta_{pp'}$ satisfying the normalization $\sum_p c_p^2=1$; ($\delta_{pp'}=0$ for $p\neq p'$, and $\delta_{pp'}=1$ for $p=p'$). The linearly combined dynamic displacement course expressed in (2) endorses the stationary action principle applied to nonlocally governed electrodynamic motion of point-like charge as $\int_{\tau_1}^{\tau_2}(\delta L)d\tau=0$, where $L\equiv -m_0c\left(\sum_p c_p^2(g_{\alpha\beta}dx_p^\alpha/d\tau)(dx_p^\beta/d\tau)\right)^{1/2} -(q/c)\sum_p c_p^2 A_\beta dx_p^\beta/d\tau$, is the Lagrangian corresponding to the weighted average path of optimum displacement course; time-like elemental displacement being, $d\tau\equiv\left(\sum_p c_p^2 g_{\alpha\beta}dx_p^\alpha dx_p^\beta\right)^{1/2}$. For the family of paths $\{x_p^\alpha\}$, the variation of Lagrangian $L \equiv \sum_p L_p(x_p^\alpha, v_p^\alpha)$ at an instant is expressible as $\delta L = \sum_p c_p^2 \delta L_p$, with $\delta L_p \equiv [(\partial L_p / \partial x_p^\mu)\delta x_p^\mu + (\partial L_p / \partial v_p^\mu)\delta v_p^\mu]$, $(v_p^\alpha = dx_p^\alpha / d\tau)$. Noting that $\delta L$ expression involves displacement components, $\{v_p^\alpha\}$, of the nonlocally correlated path family contributing to the optimized dynamic course as described by the weighted average displacement, $v^\alpha \equiv \sum_p c_p v_p^\alpha$, the stated action principle can be written in flat space ($\partial g^{\alpha\beta} / \partial x^\mu = 0$) as

$$\int_{\tau_1}^{\tau_2} \left[ \sum_p c_p^2 (\partial L_p / \partial x_p^\mu + d\pi_{\mu,p} / d\tau) \delta x_p^\mu \right] d\tau - \left[ \sum_p c_p^2 (\pi_{\mu,p} \delta x_p^\mu) \right] \Bigg|_{\tau_1}^{\tau_2} = 0 \qquad \text{(2a)},$$

where, $\pi_{\alpha,p} \equiv -(\partial L_p / \partial v_p^\alpha)$, is the canonical 4-momentum concerned with path $p$ at instant $\tau$. The dynamic course is associated with coherently evolved 4-momentum, $\pi^\alpha = \sum_p c_p \pi_p^\alpha$, which for electrodynamics is expressed as $\pi^\alpha = m_0 c v^\alpha + qA^\alpha / c$. In order to arrive at the dynamic evolution that endorses optimum displacement course, $v^\alpha \equiv \sum_p c_p v_p^\alpha$, irrespective of the arbitrarily selected variation boundaries, the ubiquitously mediating correlation necessarily implicates null value of the integrated term, that is, the second term in eq.(2a). Noting that the integrated term is essentially the difference, $(\pi_\alpha \delta x^\alpha)_{\tau_2} - (\pi_\alpha \delta x^\alpha)_{\tau_1}$ one writes the criterion for the nonlocally mediated dynamic evolution as

$$(\pi_\alpha \delta x^\alpha)_{\tau_1} = (\pi_\alpha \delta x^\alpha)_{\tau_2} = C' \text{ (say), for all } \tau_1 \text{ and } \tau_2 \qquad \text{(2b).}$$

Eq.(2b) being equally applicable even when one of the boundary values of canonical 4-momentum corroborates to that of free particle evolution under vanishingly small interacting field with the particle, then the scalar product, $(\pi_\alpha \delta x^\alpha)$ at this boundary, say, at $\tau_1$ corresponds to the value of $(m_0 c v_\alpha \delta x^\alpha)$; (for free particle, $\pi_\alpha = m_0 c v_\alpha$). Now, as discussed in the case of geodesics of quantum space, the nonlocal mediation property in the field free evolution of particles follows as that of the freely evolved geodesics. In both the cases, the scalar product, $(v_\alpha \delta x^\alpha)$ are having null values. Thus, from (2b) one concludes null value of the parameter $C'$ for all $\tau = \tau_2$. This leads to the nonlocal mediation characteristics of interacting field and particle as $\pi_\beta i^\beta = 0$, for all $\tau$, where, $i^\beta \equiv (\delta x^\beta / \delta\tau')$, with $\delta\tau' = \sqrt{-[(\delta x^0)^2 - \sum_j (\delta x^j)^2]}$, ($j = 1,2,3$), endorses the space-like 4-vector property, $i^\beta i_\beta = -1$. In other words, the canonically evolved 4-momentum, $\pi^\alpha$ maintains the 4-orthogonal connection with mediation 4-velocity, $w^{\beta} = \ell[1, \bar{\mathrm{w}}/c]$, ($\ell$ being a parameter) as

$$\pi_{\beta} i^{\beta} = \pi_{\beta} w^{\beta} = 0, \qquad \text{(2b')}$$

where, $\bar{\mathrm{w}} \equiv c(\bar{\mathrm{i}}/\mathrm{i}^0)$. For electrodynamic evolution, $\pi^{\alpha}$ is expressible as $\pi^{\alpha} = m_0 c v^{\alpha} + qA^{\alpha}/c$. As before, $\mathrm{w}$ corresponds to the nonlocal mediation speed in the virtual exchange of energy-momentum across instant evolution volume. The coherent dynamic evolution that follows from (2a) irrespective of the space-like variation $\delta\tau'$, ($\delta x^{\alpha} = i^{\alpha}\delta\tau'$), at an instant ($\tau$) is expressible as

$$(\partial L/\partial x^{\mu} + d\pi_{\mu}/d\tau)i^{\mu} = (\partial L/\partial x^{\mu} + d\pi_{\mu}/d\tau)w^{\mu} = 0 \qquad \text{(2c).}$$

The coherent evolution for electrodynamics in particular is thus expressed by

$$[\dot{v}_{\alpha} - (q/m_0 c)F_{\alpha\beta}v^{\beta}]i^{\alpha} = 0, \; (F_{\mu\nu}(x) = \partial_{\mu}A_{\nu} - \partial_{\nu}A_{\mu}) \qquad \text{(3).}$$

The scalar product $\dot{v}_{\alpha}i^{\alpha}$ in eq.(3) can be rewritten as $d(v_{\alpha}i^{\alpha})/d\tau$ because of displacement independent $i^{\alpha}$. Again, the product, $(v_{\alpha}i^{\alpha}) \equiv v_{\otimes}$(say) with $v_{\otimes} = (v_{\alpha}i^{\alpha}) = i^0_{\text{proper}}$, being independent of displacement, the product $\dot{v}_{\alpha}i^{\alpha}$ is a null. Going by the same argument, $i^{\alpha}$ shows 4-orthogonal connection with all higher derivatives of 4-velocity, which collectively convey that the nonlocally entangled evolution, as expressed for the electrodynamic case in (3), forbids involvements of observable kinematic properties. Eq.(3) thus takes the simpler form:

$$[F_{\alpha\beta}v^{\beta}]i^{\alpha} = [F_{\alpha\beta}v^{\beta}]w^{\alpha} = 0 \qquad \text{(3a).}$$

Nonlocal property, $w^{\alpha}$, or, $i^{\alpha}$ do not share components with the kinematic features like acceleration and its time derivatives all of which are the classical manifests of perturbation under external influence. Coherently interacting electromagnetic field ($F_{\alpha\beta}$) involved in coherent evolution (3a) inherently includes the contributions from omnipresent fluctuating background vacuum field (real vacuum) of cosmic space. Eq.(3a) essentially show that coherent evolution solely refers to virtual process in instantly mediated field-particle interaction, where dynamic displacement components $v^{\beta}$ are virtual in nature.

Now, as discussed in the case of coherently evolved geodesics (refer subsection A1.1), one compares the mediation characteristics as $\pi_{\beta}w^{\beta} = \sum_k c_k \pi_{k,\beta} w_k^{\beta} = 0$, with the perennial 4-orthogonal connection of nonlocal 4-velocity ($w^{\beta} \equiv \ell[1, \bar{\mathrm{w}}/c]$, with $\bar{\mathrm{w}} = \omega\bar{\mathrm{n}}_{\mathrm{k}}/\mathrm{k}$ as phase velocity) with the spectral 4-coordinates, namely, $k^{\beta}w_{\mathrm{k},\beta} \equiv 0$, $k^{\beta} \equiv [\omega/c, \bar{\mathrm{k}}]$, and infers that the nonlocally mediated canonical 4-momentum has the evolution property of $\pi_k^{\beta} \propto k^{\beta}$. The proportional connection endorses quantized evolution of the canonical 4-momentum irrespective of the interacting field and particle as $\pi_k^{\beta} = \hbar k^{\beta}$, in which the proportionality constant refers to the universal action unit, the reduced Planck constant, $\hbar = h/2\pi$.

For charged particle in electromagnetic field, one thus refers to the evolution property, $\pi_k^\beta \equiv m_0 c v_k^\beta + (q/c)A^\beta = \hbar k^\beta$, where $v^\beta = \sum_k c_k v_k^\beta$. Using the normalizations, $\sum_k c_k^2 (v_k^\alpha v_{\alpha,k}) = v^\alpha v_\alpha = 1$, and $\sum_k c_k^2 = 1$, the dynamic evolution in its scalar representation corresponds to the wave dispersion characteristics as $[\hbar k^\beta - (q/c)A^\beta]^2 = m_0^2 c^2$. In the absence of interacting field, the wave dispersion corresponds to that of freely evolving particle (point-like charge) as $[k^\beta k_\beta]^2 \equiv (\omega^2 - c^2 \mathrm{k}^2) = (2\pi c / \lambda\!\!\!^{-}_{\mathrm{m}})^2$, with $\lambda\!\!\!^{-}_{\mathrm{m}} = h / m_0 c$, as the matter-wave length. This dispersion characteristics when compared with that of coherently evolving geodesics of quantum space, namely, $(\omega^2 - c^2 \mathrm{k}^2) = (2\pi c / \lambda\!\!\!^{-})^2$, ($\lambda\!\!\!^{-}$ being basic length element of virtual lattice-like background vacuum of cosmic space), one notes that the matter wave will be resonantly supported by the space when $\lambda\!\!\!^{-}_{\mathrm{m}}$ closely matches with integral values of the basic dimension $\lambda\!\!\!^{-}$, which is shown to be of the order of Planck's length. Stability/quasi- stability of a particle is decided by the extent of matching of the ratio $\lambda\!\!\!^{-}_{\mathrm{m}} / \lambda\!\!\!^{-}$ with integral numbers.

B1.1 <u>Classical correspondence of quantum evolution –mitigatng role of background vacuum field in the course for classical transition</u>

Field concerned with the description of interacting field-particle system once externally perturbed, the nonlocal correlation does not disappear immediately by the perturbation because of ubiquitous defense safeguard from the unaltered background field associated with the virtual lattice-like gravitating quantum space of cosmos. The ubiquitous background field incessantly reacts to external perturbation to keep up nonlocally mediated instant correlation within the coherently evolved system. Sustained mitigation reaction in the wake of externally driven field leads to a brief onset course of increasingly high fluctuation in the evolution harmonics until the course culminates in thoroughly beaten state of harmonics loosing access of the nonlocal commands of the background field. Until encountering the perturbation criticality the ubiquitous field mitigation being sustained all through, the brief course culminates with optimum relaxation loss. In electrodynamic evolution of point charge in external field, two notable markers of the decoherence criticality are the acceleration kinematics and radiation relaxation as the observables in dynamics. The external power source bears the expense of relaxation loss occurring at radiation boundary. Relaxation signifies stress yield as the external drive outwits the ubiquitous defense of cosmic background field by frustrating the coherently evolved harmonics in loosing access to the nonlocal safeguard. All along the externally driven dynamic course the incessant virtual mitigation on external perturbation reveals itself as relaxation loss in successions, each following a brief onset period of about $q^2 / m_0 c^3$~$10^{-23}$s (for an electron) which is defined by radiation boundary of the moving charge. Radiation recoil from momentum loss at the boundary provides measure of the critically mitigating defense reaction in decoherence events of the dynamic course.

In perturbation driven brief onset course, the virtual mitigation reaction of ubiquitous background field will contribute additional dynamic term, $-\xi X_{\mu}(\tau)$ over the inertial and external field related terms of eq.(3). Until encountering perturbation criticality in the onset course, the nonlocally mitigated dynamics which is being increasingly perturbed is described by $[\dot{v}_{\alpha}-(q/m_0c)F_{\alpha\beta}v^{\beta}-\xi X_{\alpha}(\tau)]i^{\alpha}=0$. Virtually evolved properties with their progressively developing fluctuations constitute the dynamic terms within the square bracket. As already noted, the nonlocal 4-vector $i^{\alpha}$ is 4-orthogonal with 4-acceleration and its derivatives, and also with the field-particle interaction, $(q/m_0c)F_{\alpha\beta}v^{\beta}$ as shown in equation (3a). In view of this, 4-vector $i^{\alpha}$ is also 4-orthogonal to the mitigation reaction shown as the third term, that is, $X_{\mu}i^{\mu}=0$. Now, it is to be recalled that the nonlocally entangled description has been arrived through variation optimization of ubiquitously governed displacement of a family of paths $f(x_p^{\alpha})$ for which the displacement components and related kinematic properties are considered in the presence of nonlocal correlation existing among the path properties. Thus, the displacement components ($v^{\alpha}$) are described by involving the correlations parameters, $\{c_p\}$ as $dx^{\alpha}/d\tau=v^{\alpha}\equiv\sum_p c_p v_p^{\alpha}$, with $c_p c_{p'}=\delta_{pp'}$ ($\delta_{pp'}=0$ for $p\neq p'$, and $\delta_{pp'}=1$ for $p=p'$). The externally driven brief onset course in the presence of the virtual mitigation reaction will be reflected with abrupt increase of randomness of instantly governed nonlocal correlation parameters, $\{c_p\}$ of the path family involved in the variation optimized dynamic evolution. The set of correlations, $\{c_p\}$ so far referred by the displacement parameter, $\tau$, are now additionally influenced by the externally introduced stochastic factor which results in the increasingly high random fluctuations amidst nonlocal mitigation. In the wake of randomness, field mitigation maintains canonically averaged optimal displacement in the brief onset course. Randomly varying $\{c_p\}$ involved in the description of space-like 4-vector, $i^{\alpha}$ as $i^{\alpha}=\delta x^{\alpha}/\delta\tau'$ with $\delta x^{\alpha}=\sum_p c_p\delta x_p^{\alpha}$ results in the increasingly perturbed mediation maintaining displacement optimized in canonically averaged form over instantly chosen evolution volume in the onset course. Until the external drive could completely deprive the evolution of nonlocal safeguard the mitigation keeps up the averaged displacement which is satisfied by the canonically averaged null value of the entangled quantity, $\left\langle\left\{\dot{v}_{\alpha}-(q/m_0c)F_{\alpha\beta}v^{\beta}-\xi X_{\alpha}(\tau)\right\}i^{\alpha}\right\rangle$. The nonlocally mitigated brief onset course thus culminates with failure of the entanglement because of total arbitrariness of the nonlocal mediation feature, $i^{\alpha}$ as well as variation, $\delta x^{\alpha}$. There, the entangled description is equivalent to the equality, $\left\langle \dot{v}_\alpha - (q/m_0c)F_{\alpha\beta}v^\beta - \xi X_\alpha(\tau) \right\rangle \delta x^\alpha = 0$. This equality being valid irrespective of the arbitrary variation $\delta x^\alpha$, one gets the dynamics in classical form with the loss of nonlocal governance as

$$\left\langle \dot{v}_\alpha - (q/m_0c)F_{\alpha\beta}v^\beta - \xi X_\alpha(\tau) \right\rangle = 0 \qquad \text{(3b).}$$

Averaged dynamic terms involved in eq.(3b) are all classical quantities, which in principle can be observed. In (3b), the mitigating reaction of the background field conforms to the space-like 4-orthogonal connection, $\left\langle X_\mu(\tau)v^\mu \right\rangle = 0$. Furthermore, it is to be recalled that eq.(3b) corroborates to the null scalar product, $\left\langle X_\mu(\tau)i^\mu \right\rangle = 0$. In light of these two 4-orthogonal connections, namely, $\left\langle X_\mu(\tau)v^\mu \right\rangle = 0$, and $\left\langle X_\mu(\tau)i^\mu \right\rangle = 0$ the functional nature of $X^\alpha(\tau)$ will be discussed in the following subsection.

B1.1.1 Analysis on radiation reaction in electrodynamics

In classically referred Lorentz-Abraham-Dirac's electrodynamics [14] the radiation reaction on accelerated point-like charge was arrived with detailed consideration of energy momentum balance in the radiative motion as expressed by $-(\ddot{v}^\alpha + \dot{v}^2 v^\alpha)2q^2/3c$, ($\dot{v}^2 = \dot{v}^\alpha \dot{v}_\alpha$). The classically arrived detailed balance in externally driven dynamics inherently falls short of the consideration regarding ubiquitous involvement of nonlocal safeguard in mitigating the relaxation loss at radiation boundary referred in the dynamic course. Under the ubiquitous influence of the incessantly reacting nonlocal safeguard, the canonically averaged dynamic course encounters relaxation loss in micro-steps of optimal displacement each involving a time period of about $q^2/m_0c^3$~$10^{-23}$s, which is the brief onset course being described by eq.(3b). In this context, it is to be noted that the classically arrived LAD's radiation reaction term though conforms with the space-like property namely, $(\ddot{v}^\alpha + \dot{v}^2 v^\alpha)v_\alpha = 0$, does not conform with the mitigation feature, $\left\langle X_\mu i^\mu \right\rangle = 0$ which is evident from finite value of the scalar product, $(\ddot{v}^\alpha + \dot{v}^2 v^\alpha)i_\alpha = \dot{v}^2 v_\otimes$. The product is nonzero in accelerated dynamics ($\dot{v}^2 \neq 0$) as the nonlocal factor, $v_\otimes = i^\mu v_\mu$ in the product is an omnipresent feature which is finite. Nevertheless, it is to be noted that LAD's radiation reaction term when modified with mitigation related feature as $-(2q^2/3c)\left\langle (\ddot{v}^\alpha + \dot{v}^2 v^\alpha + \Theta i'^\alpha) \right\rangle \equiv -dR^\alpha_{\text{Total}}/d\tau$ (say), $-dR^\alpha_{\text{Total}}/d\tau$ would corroborate to the required property of nonlocal mitigation, namely, $\left\langle -(dR^\alpha_{\text{Total}}/d\tau)i^\mu \right\rangle = 0$, when $i'^\alpha = i^\alpha - v_\otimes v^\alpha$, ($i^\alpha i_\alpha = -1$), and $\Theta = \dot{v}^2[v_\otimes/(1+v_\otimes^2)]$. This can be verified by using the equalities: $i^\alpha \ddot{v}_\alpha = 0$, $\dot{v}^2 v^\alpha i_\alpha = \dot{v}^2 v_\otimes$ and $\Theta i'^\alpha i_\alpha = -\dot{v}^2 v_\otimes$. $-dR^\alpha_{\text{Total}}/d\tau$ also conforms with the space-like property: $v_\alpha dR^\alpha_{\text{Total}}/d\tau = 0$. Now, with the added modification, $-(2q^2/3c)\Theta i'^\alpha$ it is to be seen whether such modified term reveals itself as recoil effect due to momentum loss in radiation relaxation that marks the decoherence criticality. As follows, the analysis of the modified term does reveal so.

In the virtually mitigated onset of radiation relaxation, the modified term $-(2q^2/3c)\Theta i'^{\alpha}$ can be analyzed in instant commoving proper frame of the accelerated charged particle, where one writes the equalities: $\Theta i'^{\alpha} = \Theta i^{\alpha}$, and $v_{\otimes} = v_{\alpha} i^{\alpha} = \mathrm{i}^0$ with $(\mathrm{i}^0)^2 - \mathrm{i}^2 = -1$. In the frame, the factor $[v_{\otimes}/(1+v_{\otimes}^2)]$ involved in $\Theta$ works out as the ratio of $\mathrm{i}^0/[1+(\mathrm{i}^0)^2] = \mathrm{i}^0/\mathrm{i}^2$, where, $\mathrm{i}^0/\mathrm{i}^2 = (c/\mathrm{w})^2/\mathrm{i}^0$. (As elaborated already, $(i^{\alpha}/\mathrm{i}^0) = [1, \bar{\mathrm{w}}/c]$, $\bar{\mathrm{w}}$ being phase velocity with $\bar{\mathrm{w}}/c = \bar{\mathrm{i}}/\mathrm{i}^0$). Thus, in the instant proper frame one writes, $\Theta i'^{\alpha} = \dot{v}^2(c^2/\mathrm{w}^2)(i^{\alpha}/\mathrm{i}^0) \equiv \dot{v}^2(c/\mathrm{w})^2[1, (\mathrm{w}/c)\bar{\mathrm{n}}_{\mathrm{w}}]$, $\bar{\mathrm{n}}_{\mathrm{w}}$ being unit vector of the phase velocity. Noting that in the local frame, $\dot{v}^2 = \dot{v}^{\alpha}\dot{v}_{\alpha} = (\dot{\mathrm{v}}/c^2)^2$, the evolved space-like 4-vector, $-(2q^2/3c)\Theta i'^{\alpha}$ describing virtual mitigation reaction in (3b) takes the form of $-(2q^2\dot{\mathrm{v}}^2/3c^5)(c/\mathrm{w})^2[1, (\mathrm{w}/c)\bar{\mathrm{n}}_{\mathrm{w}}] \equiv -dR^{\alpha}/d\tau$ (say), $R^{\alpha} = [\mathrm{R}^0, \bar{\mathrm{R}}]$, and $\bar{\mathrm{n}}_{\mathrm{w}} \bullet \bar{\mathrm{n}}_{\mathrm{w}} = 1$. Thus, as the modified form of LAD's radiation reaction can be rewritten as $-\left\langle (2q^2/3c)(\ddot{v}^{\alpha} + \dot{v}^2 v^{\alpha}) + dR^{\alpha}/d\tau \right\rangle \equiv -dR^{\alpha}_{\mathrm{Total}}/d\tau$. Modification, $-dR^{\alpha}/d\tau$ indeed represents the ubiquitous mitigation reaction $-\xi X_{\mu}(\tau)$ of eq.(3b). In the instant co-moving proper frame, the mitigation reaction having the form of $-\mathrm{d}\bar{\mathrm{R}}/\mathrm{dt} = -\dot{\bar{\mathrm{R}}}(\mathrm{say}) = -(2q^2\dot{\mathrm{v}}^2/3c^4)(c/\mathrm{w})\bar{\mathrm{n}}_{\mathrm{w}}$, shows its reciprocal connection with magnitude of phase velocity ($\mathrm{w}$). The reaction thus conforms to null value expected for nonlocally mediated coherent evolution having infinite phase velocity for instant phase communication across evolution volume. With the inception of external perturbation, ubiquitous mitigation reaction with the null value in unperturbed state evolves out in the brief onset course and culminates with highest reaction by attaining the lowest possible value of phase velocity, ($\mathrm{w} = c$), where the highest reaction ($-\dot{\bar{\mathrm{R}}}$) is having the form of $-\dot{\bar{\mathrm{R}}}_{\mathrm{max}} = -(2q^2\dot{\mathrm{v}}^2/3c^4)[\lim_{\mathrm{w}\to c}\bar{\mathrm{n}}_{\mathrm{w}}]$. The virtual mitigation reaction is then revealed as radiation recoil due to the rate of momentum loss in the relaxation as $2q^2\dot{\mathrm{v}}^2\bar{\mathrm{n}}_{\mathrm{rad}}/3c^4$, with $\bar{\mathrm{n}}_{\mathrm{rad}}$ as unit radiation vector, ($\bar{\mathrm{n}}_{\mathrm{rad}} \bullet \bar{\mathrm{n}}_{\mathrm{rad}} = 1$). Nonlocal mediation speed is compromised down to that of relativistic limit to impart the highest mitigation effect on externally driven dynamics, thus causing minimum deviation from nonlocally governed displacement course while culminating with the loss of ubiquitous governance in external perturbation. The ubiquitous background vacuum field thus provides maximum nonlocal safeguard against external perturbation with an optimum power loss in the defense stress yield with compromised mediation speed, $\mathrm{w} = c$; yield stress propagates out of radiation boundary in the fasted way through mitigating vacuum field, and there all causality aspects are globally maintained. With the added modification, $-(2q^2/3c)\Theta i'^{\alpha}$ to LAD's radiation reaction term, one thus closely define decoherence criticality as causally connected features in the radiative dynamics.

To describe dynamics at decoherence representing culmination of virtually mitigated course of nonlocal defense, one considers eq.(3b) with the replacement of $-\xi X^{\alpha}(\tau)$ with $-dR^{\alpha}_{\text{Total}}/d\tau$. Eq.(3b) leads to the dynamic description in the local frame as $\left\langle \dot{\bar{\mathrm{v}}} - (q/m_0)\bar{\mathrm{E}} - (\mathrm{t_c}\ddot{\bar{\mathrm{v}}} + \dot{\bar{\mathrm{R}}}_{\max}) \right\rangle = 0$, ($\mathrm{t_c} = 2q^2/3m_0c^3$), where, all the dynamic terms correspond to their classical analogs; bracketed last term, $-(\mathrm{t_c}\ddot{\bar{\mathrm{v}}} + \dot{\bar{\mathrm{R}}}_{\max})$ representing modified radiation reaction in electrodynamics. $\bar{\mathrm{E}}$ is applied electric field which represents the influencing field components of $F_{\alpha\beta}$ in the instant commoving proper frame. The ubiquitous mitigation reaction term, $\left\langle -\dot{\bar{\mathrm{R}}}_{\max} \right\rangle$ being causally connected with recoil from the rate of momentum loss in radiation relaxation, one replaces it with $-\left\langle (\dot{\mathrm{v}}^2\mathrm{t_c}/c)\bar{\mathrm{n}}_{\text{rad}} \right\rangle$. Thus, dynamic description in the local frame is rewritten as

$$\left\langle \dot{\bar{\mathrm{v}}} - (q/m_0)\bar{\mathrm{E}} - \mathrm{t_c}\ddot{\bar{\mathrm{v}}} - (\mathrm{t_c}\dot{\mathrm{v}}^2/c)\bar{\mathrm{n}}_{\text{rad}}] \right\rangle = 0 \qquad (3c).$$

The 3-jerk related term in (3c) is associated with normal $\bar{\mathrm{e}}$ and binormal $\dot{\bar{\mathrm{e}}}$ components as $\ddot{\bar{\mathrm{v}}} \equiv \mathrm{d}(\dot{\mathrm{v}}\bar{\mathrm{e}})/\mathrm{dt} = (\ddot{\bar{\mathrm{v}}} \bullet \bar{\mathrm{e}})\bar{\mathrm{e}} + \dot{\mathrm{v}}\dot{\bar{\mathrm{e}}}$, ($\bar{\mathrm{e}} \bullet \dot{\bar{\mathrm{e}}} = 0$). (Correspondingly, in the four dimensional representation, 4-jerk is composed of normal and binormal 4-vectors as $\ddot{v}^{\alpha} = -(\ddot{v}^{\beta}e_{\beta})e^{\alpha} + \sqrt{-\dot{v}^2}\,\dot{e}^{\alpha}$). The radiation recoil related last term in (3c) is accordingly decomposed into the normal component, $-(\mathrm{t_c}\dot{\mathrm{v}}^2/c)(\bar{\mathrm{n}}_{\text{rad}} \bullet \bar{\mathrm{e}})\bar{\mathrm{e}}$, and binormal (tangential) component, $-(\mathrm{t_c}\dot{\mathrm{v}}^2/c)(\bar{\mathrm{n}}_{\text{rad}} \bullet \bar{\varepsilon})\bar{\varepsilon}$, ($\bar{\varepsilon} = \dot{\bar{\mathrm{e}}}/\dot{\mathrm{e}}$) and this leads to two component equations of (3c) as

$$\left\langle \dot{\bar{\mathrm{v}}} - (q/m_0)\bar{\mathrm{E}} - \mathrm{t_c}(\ddot{\bar{\mathrm{v}}} \bullet \bar{\mathrm{e}})\bar{\mathrm{e}} - (\mathrm{t_c}\dot{\mathrm{v}}^2/c)(\bar{\mathrm{n}}_{\text{rad}} \bullet \bar{\mathrm{e}})\bar{\mathrm{e}} \right\rangle = 0 \qquad (3c'),$$

and, $$-\left\langle \mathrm{t_c}\dot{\mathrm{v}}\dot{\bar{\mathrm{e}}} + (\mathrm{t_c}\dot{\mathrm{v}}^2/c)(\bar{\mathrm{n}}_{\text{rad}} \bullet \bar{\varepsilon})\bar{\varepsilon} \right\rangle = 0 \qquad (3c'').$$

Eq.(3c') can be solved for the special situation, namely, under the initial condition of $\dot{\mathrm{v}} = \dot{\mathrm{v}}_0$ at $\mathrm{t} = 0$, the external field is abruptly switched off ($\bar{\mathrm{E}} = 0$). Then, eq.(3c') yields evolution characteristics of acceleration as

$$\dot{\mathrm{v}} = \frac{\dot{\mathrm{v}}_0 \mathrm{Exp}(\mathrm{t/t_c})}{1 + (\dot{\mathrm{v}}_0/c)\int_0^{\mathrm{t}} [(\bar{\mathrm{n}}_{\text{rad}} \bullet \bar{\mathrm{e}})]\mathrm{Exp}(\mathrm{t'/t_c})\mathrm{dt'}}$$

Besides endorsing the initial condition (at $\mathrm{t} = 0$, $\dot{\mathrm{v}} = \dot{\mathrm{v}}_0$), arrived solution shows deviated evolution of acceleration value, $\dot{\mathrm{v}}$ from the runaway exponential growth with time. This is because of the presence of integral expression in the denominator where integral term arising from radiation recoil, gives impediment to the exponential growth. Ubiquitous mitigation reaction at decoherence criticality as revealed in the form of radiation recoil indirectly plays as safeguard against the runaway situation. Noting that $(\bar{\mathrm{n}}_{\text{rad}} \bullet \bar{\mathrm{e}}) \leq 1$, the general solution leads to the limiting expression of acceleration for $\mathrm{t} > 0$ as $\dot{\mathrm{v}} \geq \dot{\mathrm{v}}_0 \mathrm{Exp}(\mathrm{t}/\mathrm{t}_c)/[1+(\dot{\mathrm{v}}_0 \mathrm{t}_c/c)\{\mathrm{Exp}(\mathrm{t}/\mathrm{t}_c)-1\}]$. According to this expression, following the abrupt withdrawal of external field acceleration corresponds to the limiting values: $\underset{t\to\infty}{\mathrm{limit}}\,\dot{\mathrm{v}} \geq (c/\mathrm{t}_c)$, where the quantity, $c/\mathrm{t}_c = (2q^2/3m_0c^4)^{-1}$ represents the intrinsic acceleration for a point charge, (~$10^{31}$ms$^{-2}$ for an electron) within the dynamically relevant dimension $2\mathrm{r}_c$, ($\mathrm{r}_c \sim 2q^2/3m_0c^2 = c\mathrm{t}_c$, say) of spherically polarized vacuum field. The charge inherits acceleration, $(2q^2/3m_0c^4)^{-1}$ as a time evolved internal property of oscillation harmonics of frequency, $\omega_0 \sim 2\pi/(4\mathrm{t}_c)$. One, therefore, finds that the assumed initial value as well as the limiting values expressed by $\underset{t\to\infty}{\mathrm{limit}}\,\dot{\mathrm{v}} \geq (c/\mathrm{t}_c)$ includes the inherited acceleration. In the local frame, $\dot{\mathrm{v}}_{\mathrm{obs}}$ at $\mathrm{t}=\infty$ to be written as $\dot{\mathrm{v}}_{\mathrm{obs}} = \dot{\mathrm{v}}_{t\to\infty} - c/\mathrm{t}_c$ will tend to null value in the presence of incessantly reacting ubiquitous mitigation of background field.

Eq.( 3c") provides a measure of the binormal component of jerk related feature present in LAD's radiation reaction term. The binormal measure being $\langle\dot{\bar{\mathrm{e}}}\rangle = -(\dot{\mathrm{v}}/c)(\bar{\mathrm{n}}_{\mathrm{rad}} \bullet \bar{\varepsilon})\bar{\varepsilon}$, one finds that recoil component is opposed by manifested torque involving the angular displacement rate given by $|\langle\dot{\mathrm{e}}\rangle| \leq \dot{\mathrm{v}}/c$. In instant commoving inertial frame, observer infers that the involvement of jerk force having its binormal component, $(2q^2/3c^3)\dot{\mathrm{v}}\dot{\bar{\mathrm{e}}}$, ($\dot{\bar{\mathrm{e}}} = \mathrm{d}\bar{\mathrm{e}}/\mathrm{dt}$) is the indicator of spontaneity in decoherence where the critical mitigation reaction besides impeding on the kinematic property results in torsional drag on the polarized charged sphere within background vacuum field. The effect of torsional drag is discussed below.

B1.1.2 <u>Damping effect of the binormal jerk-component</u>

In decoherence criticality, the torsion perturbed evolution harmonics of polarized charged sphere result in damping power loss in addition to the regular relaxation loss of $2q^2\dot{\mathrm{v}}^2/3c^3 = P_{\mathrm{Larmor}}$ (say). Instant co-moving inertial observer notes that the binormal jerk component, $(2q^2\dot{\mathrm{v}}\dot{\bar{\mathrm{e}}}/3c^3)$, (with $|\langle\dot{\mathrm{e}}\rangle| \leq \dot{\mathrm{v}}/c$) registers torsion (transverse) displacement in the presence of recoil, $(\dot{\bar{\mathrm{R}}}_{\mathrm{max}} \bullet \bar{\varepsilon})\bar{\varepsilon}$ from momentum loss in regular relaxation; $\dot{\bar{\mathrm{R}}}_{\mathrm{max}} = -[(2q^2\dot{\mathrm{v}}^2\bar{\mathrm{n}}_{\mathrm{rad}})/3c^4]$. Within time $\mathrm{t}_c$ of the recoil impact, ($\mathrm{t}_c \sim \mathrm{r}_c/c$), tangential displacement being of the order of $[(\dot{\bar{\mathrm{R}}}_{\mathrm{max}}\mathrm{t}_c/m_0) \bullet \bar{\varepsilon}]\bar{\varepsilon}$, one expects that power loss due to the recoil damping to be of the order of $(2q^2\dot{\mathrm{v}}\dot{\mathrm{e}}/3c^3)(\dot{\bar{\mathrm{R}}}_{\mathrm{max}} \bullet \bar{\varepsilon})/m_0$, where, $\dot{\mathrm{e}} \leq (\dot{\mathrm{v}}/c)$. This suggests that the damping power loss has $\dot{\mathrm{v}}^4$ functional dependence as compared to the $\dot{\mathrm{v}}^2$ dependent Larmor power loss in regular radiation emission. Damping power loss can be analyzed in more details by considering the torsion driven damped harmonic evolution of internal modes of the polarized charge sphere in acceleration as briefly mentioned below. The damping power loss with quality factor of unity is expressible as $P_{\mathrm{dissip}} = [(\dot{\bar{\mathrm{R}}}_{\max} \bullet \bar{\varepsilon})^2 / (2\omega_0 m_0)]\zeta$ [15], where, frequency, $\omega_0 \sim 2\pi / (4\mathrm{t_c})$ is attributed to the three degenerate modes of spherical oscillator for which a complete oscillation involves sweeping back and forth covering the diameter, $2\mathrm{r_c}$. Factor $\zeta$ represents off-resonance frequency parameter of recoiling radiation, $\zeta = [(2\Delta\omega / \omega_0)^2 + 1]^{-1}$, where, $\Delta\omega = \omega_0 - \omega_{\mathrm{rad}}$, $\omega_{\mathrm{rad}}$ representing spectral averaged radiation frequency concerned with the regular (Larmor) relaxation spectrum, $\omega_{\mathrm{rad}} \le \omega_0$; $1/5 \le \zeta \le 1$. For the resonant oscillation frequency ($\omega_{rad} \sim \omega_0$), $\zeta = 1$. As for accelerated electron with the effective radius, $\mathrm{r_c}$, the maximum possible power loss $P_{\mathrm{dissip}}$ corresponds to blackbody temperature of the spherical oscillator surface as $(160\alpha)^{1/4}(\hbar\dot{\mathrm{v}}/2\pi c k_B) \sim 1.04(\hbar\dot{\mathrm{v}}/2\pi c k_B)$, ($\alpha = q^2 / c\hbar$ ~1/137 for electron, $k_B$ being Boltzmann constant). With the maximum possible acceleration, $\dot{\mathrm{v}}_{\max} = c^2 / \mathrm{r_c} = 3m_0 c^4 / 2q^2$, the thermal power loss as compared to the corresponding Larmor relaxation loss, is given by $P_{\mathrm{dissip}} : P_{\mathrm{Larmor}} = [(\mathrm{r_c} / c)^2 (\dot{\mathrm{v}}/c)^2_{\max}] / \pi$. At $\dot{\mathrm{v}}_{\max}$, the power ratio corresponds to $P_{\mathrm{dissip}} : P_{\mathrm{Larmor}} = 1/\pi$. This shows that with maximum possible recoil momentum corresponding to the highest accessible nonlocal defense reaction, the disordered power loss can be at the most about 24 percent of the total power loss composed of the regular and dissipative relaxations. It is interesting to recall that in the case of QV field mitigated recession of gravitationally perturbed space the two distinguishable power signatures endorsing respectively the stress accumulation and volume displacement work in the ubiquitous QV field mitigation in virtual lattice ordered quantum space of cosmic background field follow nearly the same ratio, which is of the order of 1:3 as discussed in section A1.4. In the externally driven accelerated motion of charged particle, the finite damping loss in vacuum field suggests that the dynamic reversibility is no more valid for the case. Once a dynamic evolution crosses out doorstep of the nonlocal governance, the spontaneously manifested acceleration is a kinematic property which provides measures of the regular and dissipative relaxations.

A lab observer thus concludes that charged particle aboard an accelerated frame is thermally hot and emits thermal radiation to its surrounding vacuum following the power loss functionality of $(\text{acceleration})^4$. This conclusion seemingly complement the Unruh's prediction [16] made from quantum mechanical arguments that background vacuum field described in Minkowskian space assumes thermally excited states according to an accelerated observer. Notably, Unruh predicted warmness of surrounding vacuum field is more general irrespective of whether the concerned particle aboard a non-inertial frame is electrically neutral or charged. Nevertheless, acceleration as the classically referred kinematic property is seen describing thermal state of accelerated particle with the similar functionality irrespective of the particular case of a charged specie and the general cases as per Unruh's prediction. The kinematic property as a marker of perturbation criticality governs the relaxation losses at the expense of perturbing power sources such as an electric field, or, as a gravitational field of super massive object which according to the prediction of Stephen Hawking leads to thermal radiation loss [17].

*Supplements*

S1. Evaluation of quantized space dimension ($ƛ$) from kinematic property of space recession

For the evaluation of quantized space dimension $ƛ$, consideration is made of the field equations which govern energy-momentum evolution kinematics of virtual lattice-like gravitationally perturbed cosmic space. In Friedman-Lemaitre-Robertson-Walker space the energy balance equation in space recession involving total energy density, $\rho$ and recession speed, $(\dot{a}/a)$, ($a$ being time evolved scale factor) is expressed in the general form as $(\dot{a}/a)^2 = (8\pi G/3c^2)\rho - (c^2/a^2\kappa^2)$. As this equation is applied in the limit of null matter density which corresponds ideally to the gravity field free state of flat space the last term of field equation containing curvature related quantity $\kappa$ of the time evolved 3-dimensional hyper-sphere is a null. Energy density $\rho$ in the equation corresponds to the total gravitating matters (visible +dark components) and dark energy. As indicated already, dark energy and dark matters are the two characteristic signatures involved in QV field mitigated recession of the gravitationally active virtual lattice ordered quantum space. The two signatures respectively attributed by volume displacement work and stress accumulations together account for the energy associated in field mitigated settlements of quantized cells (each of volume, $ƛ^3$) within the gravitationally active 3D virtual lattice-like network of cosmic space undergoing observable recession; energy involved for mitigated settlement of each cell being equivalent to one quantum of the field energy ($E^0_{vac}$). Thus, when the virtual lattice ordered cosmic space is ideally extrapolated to near null density of gravitating matters, $\rho$ approaches to the field energy density of $\rho^0_{vac}$ to represent space solely filled with dark energy. The ideally defined lattice-like space then corresponds to QFT referred 'real' vacuum of null space curvature holding the cosmic background field. Thus, the energy balance equation of real vacuum gets simplified to $(\dot{a}/a)^2 = (8\pi G\rho^0_{vac}/3c^2)$, where, $\rho^0_{vac} = 2\pi c\hbar/ƛ^4$. As for the equation of momentum stress balance as expressed by $(\ddot{a}/a) = -(4\pi G/3c^2)[\rho^0_{vac} + 3p^0_{vac}]$, one replaces the negative pressure $(-p^0_{vac})$ by density $\rho^0_{vac}$ according to the equation of state of repulsive QV field, namely, $p^0_{vac}/\rho^0_{xac} = -1$. Thus, the momentum conservation equation takes the form of $(\ddot{a}/a) = (8\pi G\rho^0_{vac}/3c^2)$. It shows that in the absence of gravitating matters the kinematic characteristics in idealized lattice-like space conforms to the equality of kinematic acceleration with squared kinematic velocity: $(\ddot{a}/a) = (\dot{a}/a)^2$. Kinematic property suggests of field free propagation of energy and momentum stresses with signal speed in the idealized matter-free state of lattice ordered space of background vacuum field.

As one refers the time evolved recessional properties of observable cosmic sphere, the kinematics concerned with the spherically represented virtual lattice-like quantum space of real vacuum conforms with the equalities, $(\ddot{a}/a) = (\dot{a}/a)^2 = (\dot{R}/R)^2$. $R$ as radius of the time evolved 3-D hyper-sphere which is tightly packed with the cubic cells in huge numbers. As discussed in the main text, in QV field mitigated lattice-like space, each cell of space volume, $\lambda\!\!\!^{-3}$ possesses a quantum of field energy, $E^0_{vac}$. Noting that in the idealized matter-free state ('real' vacuum) of the barrier isolated lattice-like spherical space the stress associated with the field mitigated recession propagates with signal speed, one writes $\dot{R}/R = \dot{a}/a = c/\lambda\!\!\!^{-}{}'$, where $4\pi\lambda\!\!\!^{-}{}'^3/3$ represents spherical volume equivalent to the quantized cell space, $\lambda\!\!\!^{-3}$. With the implied equality, $\lambda\!\!\!^{-}{}' = (3/4\pi)^{1/3}\lambda\!\!\!^{-}$, the energy density of background vacuum field within the cosmic sphere conforms to $\rho^0_{vac} (= 2\pi c\hbar/\lambda\!\!\!^{-4})$. The corresponding stress balance equation can be represented as $(c/\lambda\!\!\!^{-}{}')^2 = (8\pi G\rho^0_{vac}/3c^2) \equiv 16\pi^2 G\hbar/3\lambda\!\!\!^{-4}c$. This equality leads to the value of $\lambda\!\!\!^{-}$ as $\lambda\!\!\!^{-} = (4\pi/3)^{1/6}\sqrt{4\pi(G\hbar/c^3)}$ ~7.266x10$^{-35}$m, and $\lambda\!\!\!^{-}{}'$~4.51x10$^{-35}$m. Thus, $\lambda\!\!\!^{-}$ is about 4.5 times the Planck's length, $(G\hbar/c^3)^{1/2}$ ~1.616x10$^{-35}$m. It is quite clear from this analysis that Planck's length composed of the assorted physical constants is not just an empirical quantity, but it has relevance in the description of QFT referred background vacuum field of cosmic space. With the arrived $\lambda\!\!\!^{-}$ value, the kinematic speed $(\dot{a}/a)$ in critically perturbed state of the QFT referred vacuum field works out as high as $(c/\lambda\!\!\!^{-}{}')$~6.65x10$^{42}$s$^{-1}$. Thus, one finds that kinematic acceleration in critically perturbed state of the virtual lattice-like space is having ultrahigh value of $c^2/\lambda\!\!\!^{-}{}'^2$~ 4.42x10$^{85}$s$^{-2}$. Experimental probes normally used in laboratory experiments to examine quantum states of physical systems cannot perturb the background vacuum field.

in the idealized matter free space representing virtual lattice ordered cosmic background vacuum, $\lambda\!\!\!^{-}$-value arrived on the basis of observable kinematic recession need not be the exact representation of coherent length for quantum space within event horizon of a super-massive heavenly object. The lattice-like quantized background is severely perturbed under compression by ultrahigh gravity field of such object and this particularly so inside the event horizon. As the result, $\lambda\!\!\!^{-}$-value therein possibly lowers from the arrived figure of 7.266x10$^{-35}$m for the cosmic background vacuum. With an increased black hole's mass the $\lambda\!\!\!^{-}$-value can critically reach the lowest attainable level of $\lambda\!\!\!^{-}_{\min} = \sqrt{4\pi\hbar G/c^3}$ (~5.7512x10$^{-35}$m), which as indicated in the main text, represents the dimension of event horizon corresponding to quantum mass $2\pi\hbar/c\lambda\!\!\!^{-}$ attributed to the lowest mode of evolution harmonics of vacuum field. In a black hole the stated dimensional criticality would attain for an overall mass density of 2.018x10$^{95}$kgm$^{-3}$. As compared to ubiquitous QV field this density being higher by the factor of $(\lambda\!\!\!^{-}/\lambda\!\!\!^{-}_{\min})^4$,

such a dense state when attained inside a black hole will be unstable to undergo spontaneous conversion into the ubiquitous state of quantum vacuum field which is constituted of oscillator harmonics ($\omega \geq 2\pi c/ƛ$) sans the lattice-like quantum order. The critical conversion would physically corroborate to the widely believed attainment of singularity inside black hole. In the ultrahigh gravitational perturbation the excessively developed internal energy of the severely compressed lattice-like state will ease out tunneling of phonon energy across the isolation barrier to mighty QV field that is ubiquitously present. In the instability, a part of the phonon energy can take alternative tunneling course in the form of an ultrahigh energy jet out of the event horizon.

S2. Evaluation of critically developed embryonic lattice ordered kernel space

In the main text, discussion was made for the possible manifestation of ultra-hot virtual lattice ordered embryonic kernel space locally within mighty QV field through inherent fluctuation catalyzed event. Locally excited by ultrahigh fluctuation, excited QV field probabilistically can avail alternate de-excitation course involving the evanescent lattice ordered state of space in thermally ultra-hot state under barrier isolation from the precursor QV field. Once the virtual lattice ordered embryonic space is realized, field mitigation proceeds with tunneling of QV quanta across the impending energy barrier leading to spontaneous relaxation of the highly stressed state in de-excitation course under harmonic potential of the lattice. As discussed, for a critically developed kernel dimension, the mitigated relaxation course can be ultrafast through inflationary expansion with the consequential super-cooling of the initially ultra-hot embryonic lattice leading to gravitationally active space as is the observable cosmos. Now, for the estimation of critically manifested embryonic kernel's dimension, one takes the following steps:

Average thermal energy ($E_{th}$) per mode of the emergent ultra-hot kernel is evaluated by using the partition function, $Q(T) = \int_0^{E_{vac}^0} g(E)\mathrm{Exp}(-E/k_B T)dE$, $g(E) = E^2/2\pi^2 c^3 \hbar^3$ at $T = T_0$~2x10$^{32}$K, where phonon states of the virtual lattice are distributed over the energy range, $0 \leq E \leq E_{vac}^0$. Evaluation corresponds to the average energy ($E_{th}$) of a lattice-mode of about, $0.7k_B T_0$. In ultrahigh thermal state, in addition to the radiation energy content of $4\sigma T_0^4 ƛ^3/c$ per cell volume, it is assumed that the cell space accommodates phonon modes minimally with an average energy of $0.7k_B T_0$. Thermal stress of the lattice like spherical kernel space (radius, $R_k$) containing $(4\pi R_k^3/3ƛ^3) \equiv N_k$ cells (say) corresponds to total stress energy of lattice as $E_T = N_k(0.7k_B T_0)$. Accordingly, the gravitational energy content of kernel space is expressible as $-3G(E_T/c^2)^2/5R_k = E_G$ (say), where, each cell contributing average gravitation energy of ($E_G \div N_k$). Thus, the internal energy for a unit cell at $T_0$ is expressed by $E_{\text{cell}} = E_1 - G(4\pi R_k^2 / 5c^4ƛ^3)(0.7k_B T)^2$, where, $E_1 = (0.7k_B T_0 + 4\sigma T_0^4 ƛ^3 / c)$ =4.66x10$^{11}$J. This energy together with the entropy related contribution of $-k_B T_0 \ln Q_{\text{Total}}$ constitutes the internal stability of a unit cell within the embryonic lattice; $Q_{\text{Total}} = (4\pi R_{cr}^3 / 3)Q(T_0)$. For a critically developed kernel this internal stability will compensate for the energy required to surmount the isolation barrier in rendering free passage for a quantum of QV field energy for the mitigation of thermally stressed lattice. Thus, the kinematic course for barrier compensated state of the critically developed embryonic space involves energy balance equation to be expressed as $E^* + [E_{\text{cell}} - k_B T_0 \ln Q_{\text{Total}}] = 0$, for $R_k = R_{cr}$, $R_{cr}$ being critical radius. Barrier height $E^*$ is mainly governed by the distinction of quantum orders of the gravitationally perturbed lattice ordered embryonic space from that of QV. For example, with $E^* \sim 10E_{vac}^0$, the critical radius works out to be about 1x10$^{-33}$ m (~14ƛ) and correspondingly, kernel space contains about 1x10$^4$ quantized cells. With increase in height, $E^*$, the barrier compensation is critically met by larger kernel dimension; $(\partial R_{cr} / \partial E^*)$ ~ 1x10$^{-46}$ m.J$^{-1}$.

S3. Spontaneous symmetry breaking aspects in virtual lattice ordered quantum space

Coherently evolving geodesics of affine space can support additional fields over the one which is governed by space-time derivatives of the symmetric metric ($\partial_\sigma g_{\alpha\beta} / 2 \equiv \Gamma_{\sigma\alpha\beta}$). Recalling eq.(1b) of the main text one considers that the nonlocally mediated geodesic displacement property, $A_\sigma \equiv (v^\alpha v^\beta \Gamma_{\sigma\alpha\beta})$ maintains 4-orthogonal connection with mediation 4-velocity ($w^\sigma$) as $A_\beta w^\beta = 0$, and according to eq.(2b'), this 4-orthogonality can continue in the presence of additional field as $\Gamma'_{\sigma\alpha\beta} = \Gamma_{\sigma\alpha\beta} + \Pi_\sigma g_{\alpha\beta}$, where $\Pi^\mu(x)$ in the added field represents probability density of canonical 4-momentum of a scalar particle executing unitary evolution within quantum space concerned. The virtual lattice ordered affine space which is evanescently formed in ultrahigh thermal state under barrier isolation in de-excitation course of the ultrahigh fluctuation affected locality of mighty quantum vacuum field. The unitary evolution described by $[\partial_\alpha (g^{\alpha\mu} \partial_\mu) - (c\kappa_0 / \hbar)^2]\psi = 0$, involves quantum mass $\kappa_0$ of the concerned particle. Canonical 4-momentum density described by $\Pi_\mu = C(\psi^* i\hbar \partial_\mu \psi) = Cp_\mu |\psi|^2$, ($i = \sqrt{-1}$), corroborates to 4-orthogonal connection, $w^\mu \Pi_\mu = 0$ and thus contributes to the affinity $\Gamma'_{\sigma\alpha\beta}$. ($\Pi^\mu$ is expressed in inverse length unit; parameter, $C$ is having unit of inverse action). Now, with modified affinity, $\Gamma'_{\sigma\alpha\beta}$, (with $\Gamma'_{\sigma\alpha\beta} = \Gamma'_{\sigma\beta\alpha}$), the quantum space possesses symmetric pseudo space curvature, $R'_{\alpha\beta} = R_{\alpha\beta} - T_{\alpha\beta}$, ($R_{\alpha\beta}$ being Ricci curvature, $R_{\alpha\beta} = \partial_\mu \Gamma^\mu_{\alpha\beta} - \partial_\alpha \Gamma^\mu_{\beta\mu} + \Gamma^\nu_{\nu\mu}\Gamma^\mu_{\alpha\beta} - \Gamma^\mu_{\nu\alpha}\Gamma^\nu_{\mu\beta}$), where, $T_{\alpha\beta}$ is expressible as $-T_{\alpha\beta} = [-\partial_\alpha \Pi_\beta + (\partial_\mu \Pi^\mu) g_{\alpha\beta} + \Pi_\mu \partial^\mu g_{\alpha\beta}] + [(\Pi_\mu \partial^\mu \ln\sqrt{-g}) g_{\alpha\beta} + \Gamma^\mu_{\alpha\beta}\Pi_\mu - (\Gamma_{\alpha\beta\mu} + \Gamma_{\beta\alpha\mu})\Pi^\mu]$ $+[\Pi_\mu \Pi^\mu g_{\alpha\beta} - \Pi_\alpha \Pi_\beta]$. The derivative, $\partial_\alpha \Pi_\beta$ involved in $T_{\alpha\beta}$ can be explicitly written as $\partial_\alpha \Pi_\beta = (iC/\hbar)[(-i\hbar\partial_\alpha \psi^*)(i\hbar\partial_\beta \psi) + \hbar^2 \psi^* \partial_\alpha \partial_\beta(\psi)]$ and the corresponding scalar, $g_{\alpha\beta}\partial_\alpha \Pi_\beta$ is expressible as $(iC/\hbar)[p^2 + c^2\kappa_0^2]|\psi|^2$; ($i\hbar\partial_\alpha \psi = p_\alpha \psi$). Stress, $T_{\alpha\beta}$, in the pseudo curvature expression $R'_{\alpha\beta}$ is arising out of the evolved $\psi$-field. .

Now, the unitary evolution of $\psi$-field being sustained in quantum space described by affinity, $\Gamma'_{\sigma\alpha\beta}$, the overall energy-momentum stress, $[G^{\alpha\beta} + \Lambda g^{\alpha\beta} - T^{\alpha\beta}]$ is integrally conserved ($\Lambda$ being the cosmological constant). The conservation is written as $\int \partial_\beta [(G^{\alpha\beta} + \Lambda g^{\alpha\beta} - T^{\alpha\beta})\sqrt{-g}] d^3x = 0$, ($g = \det(g_{\alpha\beta})$), where, the scalar density of stress field is given by, $[g_{\alpha\beta}(G^{\alpha\beta} + \Lambda g^{\alpha\beta} - T^{\alpha\beta})\sqrt{-g}] = [-R + 4\Lambda - T]\sqrt{-g}$; $G_{\alpha\beta} g^{\alpha\beta} = -R$ and $T^{\alpha\beta} g_{\alpha\beta} = T$. By using $\psi$-evolution property one rewrites $T$ as a polynomial composed of quadratic and quartic terms of matter-wave amplitude $|\psi|$ as

$$T = -[3(iC/\hbar)(p^2 + c^2\kappa_0^2) + 7C(p^\alpha \partial_\alpha \ln\sqrt{-g})]|\psi|^2 - 3C^2[p^\mu p_\mu]|\psi|^4 .$$

In the above represented polynomial, $T$ is generally described in a complex plane of the unitary evolved $\psi$-field, which has manifested in the virtual lattice ordered affine space once such a space is evanescently formed under barrier isolation through de-excitation course of ultrahigh fluctuation affected locality of the mighty quantum vacuum. In the energy momentum conservation, the stress arising from real part of $T$ participates with those of the virtual lattice ordered affine space (namely, the scalar stress $R$ and cosmological repulsion, $\Lambda$). Real, $T$ would involve the quantum mass $\kappa_0$ of $\psi$-resonance when one considers that inverse action parameter $C$ is an imaginary quantity to be generally written as $C = \mp i/m\hbar$, ($m$ being a positive integer). In that case, the coefficient, $-3C^2$ in the last term is a positive quantity ($C^2 = -1/(m\hbar)^2$). However, the second term involving derivative of the metric determinant is the imaginary part of complex $T$, and therefore this term is not involved in the said conservation. Furthermore, one finds that with the choice of $C = -i/m\hbar$, the expression of real, $T$ turns out as Real, $T = -(3/m\hbar^2)(p^2 + c^2\kappa_0^2)|\psi|^2 + (3/m^2h^2)p^2|\psi|^4$. Scalar stress described by real, $T$ is fully symmetric with respect to $360^0$ evolved $|\psi|$ values within the complex plane. In terms of the wave-amplitude coordinates the stress shows energy minima at $|\psi| = \sqrt{[1 + (c\kappa_0/p)^2]m/2}$, which is valid for all positive integer values $m$, and these possible minima are all away from the uniquely located $T$-maximum at null value of $|\psi|$ at centre of the complex plane. The stated stress functionality $T$ corresponds to symmetry broken states that are possible from the unitary evolution related isotropic stress field in the barrier confined lattice ordered quantum space formed in evanescence. The centrally located maximum $T$ (at null $|\psi|$) corresponds to the stress of original unbroken state. For minimum potential region with a given $m$ value, the scalar-$\psi$ evolves resonantly around the central maximum of distorted potential surface with the characteristic resonance frequency, $c\kappa_0/\hbar$. As in Higgs mechanism high quantum mass corresponding to the resonance frequency once manifests in the symmetry broken state it interacts with other transiently evolving fermionic fields with the consequential manifestations of nucleonic species as the gravitating sources. As discussed in the main text, evanescent lattice ordered quantum space within barrier isolation when locally conceived in highly excited QV field under inherent fluctuation can take ultrafast relaxation course of internal stress involving the step of spontaneous symmetry breaking with consequential manifestations of the gravitating species. The intricate relaxation course involving inflationary expansion with super-cooling of the embryonic space of critically large dimension are elaborated in the text.

In order to understand the association of fermionic field along with the in-situ manifested $\psi$-resonance corresponding to high quantum mass one examines modification of the symmetric affinity, $\Gamma'_{\sigma(\alpha\beta)}$ $(=\Gamma_{\sigma(\alpha\beta)}+\Pi_\sigma g_{\alpha\beta})$ with the addition of skew symmetric field, $E_{\sigma[\alpha\beta]}=-E_{\sigma[\beta\alpha]}$. The internal degrees of freedom of fermionic fields remaining in compacted form in tensorial representation, $E_{\sigma[\alpha\beta]}$, the modified affinity, $\Gamma''_{\sigma\alpha\beta}=\Gamma'_{\sigma(\alpha\beta)}+QE_{\sigma[\alpha\beta]}$, anyway does not disturb the nonlocally correlated displacement property of quantum space: In the scalar product, $[\Gamma''_{\sigma\alpha\beta}v^\alpha v^\beta]w^\beta$, the added skew symmetric term making no contribution because of the equality, $E_{\sigma[\alpha\beta]}v^\alpha v^\beta=0$, one writes $[\Gamma''_{\sigma\alpha\beta}v^\alpha v^\beta]w^\beta=[\Gamma'_{\sigma\alpha\beta}v^\alpha v^\beta]w^\beta=0$. In the presence of fermonic field's interaction with the $\psi$-resonance of high quantum mass ($\kappa_0$) that manifested in-situ of the symmetry breaking event of high stress field associated with the barrier confined embryonic space, one finds that the modified affinity $\Gamma''_{\sigma\alpha\beta}$ leads to subtle stress features which are noteworthy. Thus, with imaginary parameter, $Q$ of the skewed affinity it is possible to arrive at a pseudo curvature involving quadratic terms in $QE_{\sigma\alpha\beta}$, which is essentially modification of the $\psi$-field affected curvature, $R'_{\alpha\beta}$, ($R'_{\alpha\beta}=R_{\alpha\beta}-\mathrm{real},T_{\alpha\beta}$). (With the use of real, $Q$, the concerned space does not approve of defining a well behaved curvature because of interference from torsion stress). With $Q^2$ as a real quantity ($Q^2<0$), the modification on $R'_{\alpha\beta}$ includes the additional terms: $Q^2[(E^\nu{}_{\mu\nu}E^\mu{}_{\alpha\beta}-E^\nu{}_{\alpha\mu}E^\mu{}_{\nu\beta})+E^\mu{}_{\alpha\nu}(E^\nu{}_{\mu\beta}-E^\nu{}_{\beta\mu})]$. Now, skewed affinity $E_{\sigma\alpha\beta}$ can be described by functionalities such as $j_o\Phi_{\alpha\beta}$ and $(j_\alpha\Phi_{\sigma\beta} - j_\beta\Phi_{\sigma\alpha})$ with $\Phi_{\alpha\beta} = -\Phi_{\beta\alpha}$, where, $j^\mu$ represents 4-current density associated with the $\psi$-resonance (quantum mass, $\kappa_0$); $j_\mu = \psi^\dagger\partial_\mu\psi - (\partial_\mu\psi^\dagger)\psi$, $(\partial_\alpha j^\alpha = 0)$. The 4-current density is involved in scattering interaction with the fermionic field $\Phi_{[\alpha\beta]}$. (Spinor field $\varphi_a$ in the compacted form involving Dirac's $\gamma_\mu$ matrices in the instantly referred Minkowskian space coordinates described by $\bar{\varphi}(\gamma_\alpha\gamma_\beta - \gamma_\beta\gamma_\alpha)\varphi$ can represent $\Phi_{[\alpha\beta]}$). Imaginary parameter $Q$ associated with the skewed affinities involves the characteristic range, $\alpha_R$ $(\alpha_R = \hbar / m_{ib}^0 c)$ of intermediate boson (charged/neutral) of mass $m_{ib}^0$ which mediates in the scattering interaction. Now, the skewed affinity of the form $QE_{\sigma\alpha\beta} = Q(j_\alpha\Phi_{\sigma\beta} - j_\beta\Phi_{\sigma\alpha})$ has a special feature to be mentioned here. Thus, one considers modification of the symmetric stress field of embryonic space due to the inclusion of the skewed affinity, $Q(j_\alpha\Phi_{\sigma\beta} - j_\beta\Phi_{\sigma\alpha})$. The modification is expressible as $Q^2[2S_\alpha j_\beta - (j_\nu\Phi^\nu{}_\alpha)(\Phi^\mu{}_\beta j_\mu) - (j_\alpha j_\beta)(\Phi^\nu{}_\mu\Phi^\mu{}_\nu)]$, where, $S_\alpha \equiv (j_\nu\Phi^\nu{}_\mu\Phi^\mu{}_\alpha)$. Among the contributed terms in this modification, the first one describing coupling of scattering complex $(S_\alpha)$ with 4-current $(j_\beta)$ can be explicitly written as $2Q^2 S_\alpha j_\beta \equiv Q^2[(S_\alpha j_\beta + S_\beta j_\alpha) + (S_\alpha j_\beta - S_\beta j_\alpha)]$, in which asymmetry in $\alpha\beta$ indices is apparent. Pseudo curvature involving the skewed stress component, $(S_\alpha j_\beta - S_\beta j_\alpha)$ can lead to biasness in manifestations of fermions of opposite charges through Higgs mechanism in symmetry breaking events that occur in the field mitigated ultrafast relaxation course entailing space inflationary super-cooling of the barrier confined virtual lattice ordered embryonic kernel as elaborated in the text.